\let\ORIlabel\label%
\let\ORIrefstepcounter\refstepcounter%
\AddToHook{package/hyperref/before}{%
   \let\label\ORIlabel%
   \let\refstepcounter\ORIrefstepcounter%
}
\documentclass[final, hidelinks, onefignum, onetabnum, nohypdvips]{siamart220329}

\usepackage{lipsum}
\usepackage{amsfonts}
\usepackage{amsopn}
\usepackage{bm}

\usepackage{cite}
\usepackage{amssymb}
\usepackage[letterpaper, margin=1.2in]{geometry}

\usepackage{graphicx}
\usepackage{epstopdf}
\usepackage{algorithmic}
\usepackage[normalem]{ulem}
\usepackage[T1]{fontenc} %
\usepackage{url} %
\usepackage{booktabs} %
\usepackage{nicefrac} %
\usepackage{microtype} %
\usepackage{lipsum}
\usepackage{fancyhdr} %
\usepackage{graphicx} %
\usepackage[textsize=scriptsize]{todonotes}
\usepackage{tikz, pgfplots, colortbl}
\usepackage{xpatch}
\usepackage{multirow}
\usepackage{hyperref} %
\usepackage[capitalize]{cleveref}

\graphicspath{{figs/}}
\pgfplotsset{compat=1.18}

\floatname{algorithm}{Algorithm}

\ifpdf%
\DeclareGraphicsExtensions{.eps,.pdf,.png,.jpg} \else \DeclareGraphicsExtensions{.eps}
\fi

\newsiamremark{remark}{Remark} \newsiamremark{hypothesis}{Hypothesis}
\crefname{hypothesis}{Hypothesis}{Hypotheses} \newsiamthm{claim}{Claim}

\headers{Sampling through iterated approximation}{D. Sharp, B. van Bloemen Waanders, Y. Marzouk}

\title{
    Sampling through iterated approximation: Gradient-free and multi-fidelity Bayesian inference via transport
}

\author{%
Daniel Sharp\thanks{Center for Computational Science and Engineering, Massachusetts Institute of Technology, Cambridge, MA (\email{dannys4@mit.edu}, \email{ymarz@mit.edu})}%
\and Bart van Bloemen Waanders\thanks{Sandia National Laboratories, Albuquerque, NM (\email{bartv@sandia.gov})}%
\and Youssef Marzouk\footnotemark[2]%
}

\makeatletter
\xpatchcmd{\algorithmic}{\itemsep\z@}{\itemsep=2pt}{}{}
\makeatother

\newif\ifcomments
\commentsfalse

\newcommand{\EE}{\mathbb{E}}
\newcommand{\RR}{\mathbb{R}}
\newcommand{\NN}{\mathcal{N}}
\newcommand{\FF}{\mathcal{F}}
\newcommand{\bfx}{\mathbf{x}}
\newcommand{\bfy}{y}

\newcommand{\bfc}{c}

\newcommand{\eps}{\varepsilon}

\newcommand{\KL}{\mathrm{KL}}

\newcommand{\PDEDomain}{\mathcal{X}}
\newcommand{\E}[1]{$\times 10^{#1}$}

\newcommand{\MMD}{\mathrm{MMD}}
\newcommand{\rESS}{\mathrm{rESS}}
\newcommand{\MIS}{\mathrm{MIS}}

\newcommand{\prior}{{\pi_{\theta}}}
\newcommand{\likely}{\pi^{\bfy^*}}
\newcommand{\poster}{{\pi}}
\newcommand{\surrogate}{\widetilde{\poster}}
\newcommand{\refer}{\eta}

\newcommand{\fidelity}{{\ell}}
\newcommand{\temper}{\beta}

\newcommand{\likelyFidel}{{\pi^{\bfy^*}_{\fidelity}}}
\newcommand{\likelyFidelP}{{\pi^{\bfy^*}_{\fidelity+1}}}

\newcommand{\stepIDX}{{j}}
\newcommand{\temperStep}{\beta_{\stepIDX}}
\newcommand{\fidelityStep}{{\ell_{\stepIDX}}}

\newcommand{\likelyStep}{{\pi^{\bfy^*}_{\fidelity_{\stepIDX}}}}
\newcommand{\posterStep}{{\poster_{\stepIDX}}}
\newcommand{\surrogateStep}{{\surrogate_{\stepIDX}}}

\newcommand{\surrQuad}{\widetilde{Q}}
\newcommand{\refweight}{\widetilde{w}}
\newcommand{\unnormw}{\nu} %

\definecolor{english}{rgb}{0.0, 0.5, 0.0}

\ifcomments
\newcommand{\ymmtext}[1]{\textcolor{english}{#1}}
\newcommand{\dgstext}[1]{\textcolor{red}{#1}}
\newcommand{\todofinal}[1]{\todo[color=black,textcolor=white]{#1}}
\newcommand{\DGS}[1]{\todo{D\#: #1}}
\newcommand{\DGSi}[1]{\todo[inline]{D\#: #1}}
\newcommand{\ymm}[1]{\todo[color=green!40]{ymm\@: #1}}
\newcommand{\ymmi}[1]{\todo[color=green!40,inline]{ymm\@: #1}}

\newcommand{\bartnote}[1]{\todo[color=green,textcolor=black]{Bart: {#1}}}
\newcommand{\bartnotei}[1]{\todo[color=green,textcolor=black,inline]{Bart: {#1}}}
\else
\renewcommand{\todo}[1]{}
\newcommand{\ymmtext}[1]{#1}
\newcommand{\dgstext}[1]{}
\newcommand{\todofinal}[1]{}
\newcommand{\DGS}[1]{}
\newcommand{\DGSi}[1]{}
\newcommand{\ymm}[1]{}
\newcommand{\ymmi}[1]{}

\newcommand{\bartnote}[1]{}
\newcommand{\bartnotei}[1]{}
\fi

\newfam\hebfam
\font\tmp=rcjhbltx at10pt \textfont\hebfam=\tmp
\font\tmp=rcjhbltx at7pt  \scriptfont\hebfam=\tmp
\font\tmp=rcjhbltx at5pt  \scriptscriptfont\hebfam=\tmp

\edef\declfam{\ifcase\hebfam
     0\or1\or2\or3\or4\or5\or6\or7\or8\or9\or A\or B\or C\or D\or E\or F\fi}

\mathchardef\shin   = "0\declfam 98 %

\ifpdf%
\hypersetup{
	pdftitle={Sampling through iterated approximation},
	pdfauthor={D. Sharp, B.v.B. Waanders, Y. Marzouk}
}
\fi

\makeatletter
\@mparswitchfalse%
\makeatother
\normalmarginpar%

\begin{document}
\maketitle

\begin{abstract}
    We develop an iterative framework for Bayesian inference problems where the posterior distribution may involve computationally intensive models, intractable gradients, significant posterior concentration, and pronounced non-Gaussianity.
Our approach integrates: (i) a generalized annealing scheme that combines geometric tempering with multi-fidelity modeling; (ii) expressive measure transport surrogates for the intermediate annealed and final target distributions, learned variationally without evaluating gradients of the target density;
and, (iii) an importance-weighting scheme to combine multiple quadrature rules, which recycles and reweighs expensive model evaluations as successive posterior approximations are built. Our scheme produces both a quadrature rule for computing posterior expectations and a transport-based approximation of the posterior from which we can easily generate independent Monte Carlo samples.
We demonstrate the efficiency and accuracy of our approach on low-dimensional but strongly non-Gaussian Bayesian inverse problems involving partial differential equations.

\end{abstract}



\section{Introduction}
\label{sec:intro}

Bayesian inference uses observational data and prior knowledge to learn or `calibrate' the parameters of a mathematical model. Commonly, a scientist or engineer prescribes a computational simulation realizing the model, i.e., mapping parameters to predicted observations, and this simulation is embedded in a {likelihood} function that represents the conditional distribution of the data. A {prior} probability distribution encodes knowledge about the parameter values; this distribution is  updated, via the likelihood, to produce a \textit{posterior distribution} on the parameters of interest. The computational task of Bayesian inference is to characterize this posterior distribution---for instance, by estimating   %
 posterior expectations of arbitrary test functions. %
Modern methods for Bayesian computation typically grapple %
with three key challenges:
First, the likelihood may be computationally expensive to query.
Second, gradients and higher-order derivatives of this likelihood may be unavailable or computationally intractable.
Third, observational data may favor narrow and geometrically complex regions of the
parameter space; in this sense, the data are informative and the likelihood is strongly
\textit{concentrated} with respect to the prior. %

In this paper, we introduce new computational Bayesian inference techniques for problems where all three of these challenges are central. Our focus is on problems where evaluations of the underlying simulation (and hence the likelihood) are
expensive, such that one can afford at most a few hundred such queries, perhaps much less.
The question of \textit{where} in the parameter space to make these queries then becomes crucial, but the answer is
complicated by the combination of posterior concentration and complex geometry.
Designing queries according to the prior ``wastes'' model evaluations
in regions of vanishing posterior probability, but placing queries in regions of high posterior probability requires not only finding these regions but describing their structure.
We must therefore develop methods that capture strong non-Gaussianity, e.g., well-separated posterior modes or sharp concentration around nonlinear ridge-like structures.
And while our black-box simulations do not permit gradient evaluations, these simulations often admit lower-fidelity approximations. Thus, we seek an inference framework that naturally exploits multiple model fidelities if they are available.

The core computational contribution of this paper is %
\ymmtext{a novel iterative approach that uses multiple model fidelities and efficient query design to effectively neutralize the core challenges of Bayesian inference described above. In particular, our approach}
combines annealing, quadrature, multiple importance sampling, and variational approximations of the posterior $\poster$. %
In a given iteration $\stepIDX$, we first choose a target distribution $\posterStep$ using tempering (less concentrated than $\poster$) and low-fidelity modeling (cheaper to evaluate than $\poster$). We then use $\surrogate_{\stepIDX-1}$, the previous iteration's target approximation, to generate new samples or quadrature points, and apply multiple importance sampling to discretize an expectation over the current target $\posterStep$. This process draws from and augments a dataset of model evaluations accumulated over previous %
iterations. The resulting discretization allows us to evaluate and minimize an objective function whose minimizer is $\surrogateStep$, a variational approximation of $\posterStep$. The iterations then continue.

In related importance sampling methods, practitioners often select simple distribution approximations (which we also call \textit{surrogates}) $\surrogateStep$, e.g., Gaussians or Gaussian mixtures, %
at the cost of high sampling bias and inefficiency; these issues are exacerbated by posterior concentration and complex geometry. %
In contrast, we use a flexible and robust \textit{measure transport} %
approach to capture non-Gaussian behavior~\cite{villaniOptimalTransportOld2009,rezende2015variational,baptistaRepresentationLearningMonotone2023,albergo2024learning,ramgraberFriendlyIntroductionTriangular2025,elmoselhyBayesianInferenceOptimal2012,marzoukSamplingMeasureTransport2016}. Moreover, our approach
builds transport-based surrogates without the gradient evaluations typically required to use transport in Bayesian inference~\cite{brennanGreedyInference2020,elmoselhyBayesianInferenceOptimal2012}.%
Additionally, rather than using only multi-fidelity modeling or only tempering, we combine the two into a \textit{generalized annealing process}. By integrating these notions of annealing with a versatile method to re-use evaluations of $\pi$, our approach makes particularly efficient use of expensive model evaluations, tailoring them to the geometric features of the posterior. %
Furthermore, our iterative scheme has a built-in robustness, described below, %
that mitigates the impact of a poor target approximation at any intermediate step. %
Synthesizing these methods yields a sequential inference approach that largely parallelizes and minimizes model evaluations, adaptively approximates the target distribution, and learns a normalized posterior density surrogate from which we sample exactly---all without gradients of the unnormalized target density.

\subsection{Related work}
\label{sec:relatedwork}

To put our contributions in context, we outline previous methods and the extent to which they address the three challenges stated in the opening paragraph.

There is of course an enormous literature on Markov chain Monte Carlo (MCMC) methods in Bayesian computation~\cite{metropolisEquationStateCalculations1953,haarioAdaptiveMetropolisAlgorithm2001,brooksHandbookMarkovChain2011,marzoukStochasticCollocation2009}.
The first MCMC methods for expensive likelihoods trained offline surrogates for the forward model, typically by minimizing a prior-weighted approximation error~\cite{marzoukStochasticSpectralMethods2007,marzoukDimensionalityReductionPolynomial2009,eldredRecentAdvancesNonIntrusive2009,marzoukStochasticCollocation2009}. While broadly effective, these methods deteriorate under high posterior concentration~\cite{lu_limitations_2015}. %
Other methods have incorporated multiple model fidelities in the proposal process, via delayed acceptance schemes or multiple interacting chains~\cite{higdonBayesianApproachCharacterizing2002,christenMarkovChainMonte2005,peherstorferSurveyMultifidelityMethods2018}. A different class of methods create and refine surrogates for the forward model or likelihood function online, i.e., as MCMC iterations proceed~\cite{conradAcceleratingAsymptoticallyExact2016,conradParallelLocalApproximation2018,zhangAcceleratingMCMCKrigingbased2019,davisRateoptimalRefinementStrategies2022} %
possibly incorporating multi-fidelity data~\cite{yanAdaptiveMultifidelityPolynomial2019}. While all of these methods may reduce the number of expensive model evaluations, they do not necessarily ameliorate the problem of geometry.
Samplers focused on challenging non-Gaussian target geometry typically require considerable information about the target density, e.g., repeated gradient and possibly Hessian evaluations~\cite{martinStochasticNewtonMCMC2012}. Alternatively, one can attempt to ``neutralize'' difficult geometry of the target in various ways. Parallel tempering and simulated annealing are widely used to mitigate concentration or promote mixing between modes. More modern methods use measure transport to transform the target distribution into a distribution that is easier to sample~\cite{parnoTransportMapAccelerated2018,albergo2024learning,gabrieAdaptiveMonteCarlo2022} or, in the opposite direction, use transport to create a biasing distribution in importance sampling (IS) or a proposal in a Metropolis independence sampler~\cite{midgleyFlowAnnealedImportance2023,mullerNeuralImportanceSampling2019,gabrieAdaptiveMonteCarlo2022,robertMonteCarloStatistical2004}.

The latter methods actually straddle the line between Monte Carlo sampling and variational approximation, as their transport maps typically are built variationally. One advantage of the variational approach is that it allows embarrassingly parallel evaluations of the target distribution, which is less natural for MCMC~\cite{souzaParallelMCMCEmbarrassing2022,neiswangerAsymptoticallyExactEmbarrassingly2014,jacobUnbiasedMarkovChain2020,calderheadGeneralConstructionParallelizing2014}. %
A fundamental problem in variational approximation, however, is the frequent use of \textit{reverse Kullback--Leibler} (KL) divergence as an optimization objective. Minimizing reverse KL requires evaluating or estimating~\cite{pmlr-v33-ranganath14} gradients of the target log-density, but more importantly, it is intrinsically \textit{mode-seeking}. Capturing posteriors that concentrate onto multiple modes or other non-Gaussian structures is therefore quite difficult.
While there are many variational approximation schemes for Bayesian inference~\cite{naessethMarkovianScoreClimbing2020,rezende2015variational,NEURIPS2022_5d087955,katsevich_approximation_2024,bleiVariationalInferenceReview2017a}, virtually all struggle with the mode-seeking nature of reverse KL. Even using tempering
for multi-modal targets does not circumvent the loss function's basic issues~\cite{zangerSequentialTransportMaps2024}. %
A few methods use alternative loss functions to mitigate mode collapse, avoid gradients, and improve stability~\cite{felardosDesigningLossesDatafree2023}. %
Ensemble Kalman samplers represent an alternative approach to gradient-free inference but have difficulty capturing non-Gaussian posteriors~\cite{garbuno-inigoInteractingLangevinDiffusions2020}.%

Perhaps the closest methods to ours are those that iteratively create a sequence of target approximations, often involving some notion of tempering. Traditionally, these have been cast as cross-entropy methods for sequential IS~\cite{rubinsteinCrossEntropyMethodCombinatorial1999,kroese_cross-entropy_2013,andrieuGradientfreeOptimizationIntegration2024,chopinIntroductionSequentialMonte2020}. %
Other recent efforts try to extend the simple biasing distributions in IS by approximating other elements of the problem, and/or by using more sophisticated approximation methods. For example,~\cite{liAdaptiveConstructionSurrogates2014} and \cite{farcasMultilevelAdaptiveSparse2020} build quadrature- or interpolation-based surrogates of the likelihood. \cite{liAdaptiveConstructionSurrogates2014} performs tempering to iteratively refine a polynomial approximation of the forward model, using quadrature over an adapted Gaussian measure, whereas \cite{farcasMultilevelAdaptiveSparse2020} interpolates on Gaussian Leja points to capture the ratios of log-likelihoods induced by models of different fidelity, without tempering.
Fundamentally, both of these efforts rely on Gaussian approximations. Moreover, they do not give a posterior surrogate that one can sample in closed-form; for instance, the algorithm of \cite{liAdaptiveConstructionSurrogates2014} culminates in MCMC sampling.
Yet both of these efforts provide useful inspiration for the present work, where we use multiple model fidelities and tempering, along with quadrature, to build posterior approximations more efficiently. We also rely on non-Gaussian variational approximations, found by minimizing \emph{forward KL} rather than reverse KL, without the evaluation of likelihood gradients.

Concurrent work~\cite{wangMitigatingModeCollapse2025,kimSequentialNeuralJoint2025}  has also introduced iterative variational approximations based on forward KL minimization and tempering. %
As we discuss later, these schemes use accumulated samples without regard to the quality of the approximation that produced them, and thus are not robust to poor iterations in the sequence.
Further, they do not emphasize sample efficiency, as demanded by expensive models; examples even in low dimensions~\cite{wangMitigatingModeCollapse2025} require at least $10^5$ evaluations of the target.
The interacting particle samplers in~\cite{mauraisSamplingUnitTime2024,wang2024measure} %
can be understood as a gradient-free tempering-based scheme, but again do not focus on parsimonious model evaluations and only return a set of samples, whereas we create %
a generative model for the entire posterior distribution. %

\subsection{Outline}
In \Cref{sec:formulation}, we formalize many of the notions discussed in
the introduction and propose a basic iteration for Bayesian inference via the cross-entropy method and generalized annealing. %
In \Cref{sec:transport}, we describe how to create a measure transport surrogate at each iteration, identifying the surrogate using importance-weighted quadrature.
In \Cref{sec:algorithms}, we detail the algorithms we implement computationally: first, a more advanced strategy of ``multiple tempered importance-weighted quadrature'' to re-use model evaluations across iterations; %
and second, our full algorithmic approach, which combines the multiple tempered quadrature with multi-fidelity modeling, tempering, measure transport, and cross-entropy minimization. We comment on several algorithmic choices in \Cref{sec:remarks}, relating these choices to the broader literature. %
Finally, in \Cref{sec:examples}, we demonstrate results on increasingly difficult problems: unimodal and bimodal posteriors for the source term of an elliptic PDE, a posterior over initial conditions for nonlinear convection-diffusion-reaction, and a highly concentrated posterior for inferring a parameter of a Helmholtz PDE.

\section{A framework for sequential Bayesian inference}
\label{sec:formulation} The Bayesian approach to statistical inference describes uncertainty in
parameters $\theta$ probabilistically, before and after observing some data $\bfy$. We work in
the setting where both $\theta$ and $\bfy$ are real vectors that have densities with respect to
Lebesgue measure. Bayes' rule can then be written in terms of density functions:
\begin{equation}
	\pi_{\theta|Y}(\theta|\bfy) = \frac{\pi_{Y|\theta}(\bfy|\theta)\prior(\theta)}{\pi_{Y}(\bfy)}\propto
	\pi_{Y|\theta}(\bfy|\theta)\prior(\theta). \label{eq:bayesrule}
\end{equation}
We consider a fixed realization $\bfy^{*}$ of the observation and seek to approximate the posterior density $\pi_{\theta|Y}(\theta|\bfy^{*})$ in~\eqref{eq:bayesrule}, for which we use the shorthand $\poster \equiv
\pi_{\theta|Y}$.

In this section, we describe an iterative framework for building probabilistic surrogates, i.e., approximations of $\pi$. %
It is rooted in sequential importance sampling, though one output of our framework is a surrogate that can be used on its own, \textit{without} importance weights. %
A key idea is the incorporation of tempering and multi-fidelity modeling into the iterations.
Before describing novelties in \cref{sec:transport,sec:algorithms},
we present the fundamental framework needed to understand them.

\subsection{Biasing distributions and posterior surrogates}
Most often, ``characterizing the posterior'' amounts to computing the expectation of some
function $h: \Omega \to \RR^{q}$, $q \geq 1$, under the posterior $\poster$. Different choices of $h$
correspond to, for example, the posterior mean, higher-order moments, or event probabilities. A simple
and gradient-free approach to estimating such expectations is importance sampling (IS): Suppose we
have identified a \textit{biasing distribution} with density $\surrogate:\Omega\to\RR^{+}$.
Then the posterior expectation of $h$ is estimated as
\begin{equation}
	\label{eqn:importance_samp}\EE_{\poster}[h] = \EE_{\surrogate}\left[h\,\frac{\,\poster\,}{\surrogate}
	\right] \approx \frac{1}{N}\sum_{i=1}^{N}h(\theta^{(i)})\ \frac{\poster(\theta^{(i)})}{\surrogate(\theta^{(i)})},
\end{equation}
where $\theta^{(i)}\stackrel{\textrm{i.i.d.}}{\sim} \surrogate$, and $\surrogate$ can be understood as a surrogate for the target
$\poster$. This Monte Carlo estimator converges almost surely to the true expectation for integrable
$h$ under mild assumptions on $\surrogate$ and $\poster$, but the variance of the IS estimator is drastically
affected by the discrepancy between $\surrogate$ and $\poster$. In particular, if $\surrogate$ assigns
significantly higher probability to certain regions of $\Omega$ compared to $\poster$, the variance
of the estimate can be large or even unbounded; similarly, if $\poster$ assigns high probability to
some region that $\surrogate$ ignores, the variance can become arbitrarily large as well.
Controlling $\poster/\surrogate$, the ratio of the two densities, is then essential to good
behavior of IS~\cite{owenMCBook2013}.
For simplicity, \Cref{eqn:importance_samp} assumes that the densities $\poster$ and $\surrogate$ are normalized. In practice,
however, the normalization constant of $\poster$ is seldom known and one must resort to a ``self-normalized'' %
version of the IS estimator (which normalizes the weights) with similar performance considerations for the biasing
distribution~\cite{owenMCBook2013}. %
Note that evaluating the IS estimator on the right of~\eqref{eqn:importance_samp} crucially requires both the
ability to draw samples from $\surrogate$ and to evaluate its (unnormalized) density. We therefore need to choose $\surrogate$ within a family of distributions $\mathcal{P}(\Omega)$ for which these properties hold.

We defer the discussion of good families $\mathcal{P}(\Omega)$ to \Cref{sec:transport} and first consider how to find $\surrogate$ within a given family. %
While the variance of the estimator in \eqref{eqn:importance_samp} can be directly related to the square of $\poster / \surrogate$, it is well established in the IS literature that a more stable objective is the log-ratio $\log \poster / \surrogate$, and hence it is useful to choose a $\surrogate$ that minimizes the \textit{forward Kullback--Leibler} (KL) divergence:
\begin{equation}
	\label{eqn:loss_minimization}\surrogate^{*} \in  \arg\min_{\surrogate\in\mathcal{P}(\Omega)}D_{\KL}(\poster
	\|\surrogate) = \arg\min_{\surrogate\in\mathcal{P}(\Omega)}\EE_{\poster}\left[\log\frac{\poster}{\surrogate}
	\right] = \arg\min_{\surrogate\in\mathcal{P}(\Omega)}\mathcal{L}[\surrogate],\quad \mathcal{L}[\surrogate]
:= -\EE_{\poster}\left[\log \surrogate\right],
\end{equation}
where $\mathcal{L}$ is the \textit{cross-entropy loss}.\footnote{The equivalence of minima in \eqref{eqn:loss_minimization} is because the entropy term $\EE_{\pi}[\log\poster ]$ is constant with respect to $\surrogate$.} %
We note that the minimizer in \eqref{eqn:loss_minimization} does not depend on the normalization of $\poster$.
By construction, and in contrast with certain IS methods, the objective in \eqref{eqn:loss_minimization} does not contain $h$, as it is intended to be robust for a wide range of posterior expectations.%

If the identified $\surrogate$ is sufficiently close to $\poster$, then it is reasonable to use expectations over $\surrogate$ without IS corrections; the bias of these unweighted estimates will be sufficiently small, and the computational savings---through avoiding evaluations of the likelihood---will be significant. %
This is one of our goals: a key output is the surrogate itself,
which allows us to draw i.i.d.\ samples from an approximate posterior.
Importance sampling, via the iterations described next, can be seen as a means toward this goal. Using an expressive family $\mathcal{P}(\Omega)$ and accurately evaluating the loss $\mathcal{L}$ will also be essential to achieving an accurate ``standalone''  $\surrogate$.

\subsection{Generalized annealing}

Evaluating the cross-entropy loss $\mathcal{L}$ requires integrating over $\poster$, which we previously assumed impossible or impractical.
We will approximate this expectation using IS, but through a sequence of simpler distributions that bridge between the simplicity of the prior and the complexity of the posterior $\poster$. We only create the surrogate for $\poster$ at the end of this sequence.
This strategy is a \textit{generalized annealing} process, incorporating both tempering and multi-fidelity modeling.

While our approach will be applicable to generic forms of the likelihood function
$\theta \mapsto \pi_{Y|\theta}(\bfy^{*}|\theta)$, we first illustrate it via a likelihood that
involves a computationally intensive model and additive Gaussian noise. Let $G:\Omega\to\RR^{m}$ be
some (generally nonlinear) function, termed the ``forward model,'' acting on parameters $\theta
\in \Omega \subseteq \RR^{d}$. The data are then modeled as $\bfy = G(\theta) + \epsilon$, where $\epsilon
\sim \mathcal{N}(0, \Sigma)$ with covariance matrix $\Sigma \in \RR^{m \times m}$. This relationship
yields the likelihood function:
\begin{equation}
	\label{eqn:gaussian_obs}\pi_{Y|\theta}(\bfy |\theta) = \mathcal{N}(\bfy;G(\theta ),\Sigma) \propto
	(\det|\Sigma|)^{-1/2}\exp\left[-\frac{1}{2}(G(\theta) - \bfy)^{\top}\Sigma^{-1}(G(\theta)-\bfy)\right
	],
\end{equation}
where the constant of proportionality is twice Archimedes' constant (i.e., $6.283 1\ldots$).\footnote{Not
to be confused with $\pi$, our notation for density.}
Since the observation $\bfy^{*}$ is fixed, we will use the shorthand
$\likely(\theta) := \pi_{Y|\theta}(\bfy^{*}|\theta)$ for the likelihood, similar to our notation
$\poster(\theta) := \pi_{\theta|Y}(\theta|\bfy^{*})$ for the posterior density.

Tempering, well-established
in sampling literature~\cite{geyerAnnealingMarkovChain1995,chopinIntroductionSequentialMonte2020}, raises
the likelihood to some fractional powers $\temper \in [0,1]$, given by an increasing sequence of \textit{inverse tempereature} parameters $(\temper_{j})_{j=0}^M$ with $\temper_M = 1$. As $j$ approaches $M$, the sequence of distributions $[\likely]^{\temper_j}\prior$ transforms
from the prior distribution ($\temper = 0$) to the posterior distribution ($\temper = 1$). Since the
prior is uninformed by data, we can generally view this sequence
as passing from minimal concentration\slash maximal
uncertainty to maximal concentration\slash minimal uncertainty. For the Gaussian noise model described
above, this effect is explicit: $[\likely(\theta)]^{\temper} = \NN(\bfy^{*}; \, G(\theta),\Sigma/ \temper
)$, and hence a smaller $\beta$ %
induces a larger covariance matrix to broaden the likelihood.

The key idea of tempered sequential IS is to use  $[\likely]^{\temper_j}\prior$ as a biasing distribution for expectations over $[\likely]^{\temper_{j+1}}\prior$. For this choice to be effective, the former should cover all the high-probability regions of the latter and hence control %
the quantity
$\log \poster/\surrogate$ discussed above. In other words, the goal is to develop points with nearly uniform (i.e., low variance) weights, taking advantage of the fact that small changes in $\beta_j$ should produce small changes in density.
The tempered IS construction
is applied iteratively. Moreover, since sampling from $[\likely]^{\temper_j}\prior$ is not directly feasible, in our approach we learn biasing distributions that \textit{approximate} the intermediate targets $[\likely]^{\temper_j}\prior$. The form of these iterations is detailed in \Cref{sec:core_method_outline}.

While tempering helps address challenges of geometry, e.g., concentration of the posterior with respect to the prior, it does not reduce the computational costs of an expensive likelihood or forward model $G$.
We will employ multi-fidelity modeling, in a similar sequential manner, to address this issue.
Using the Gaussian
setting of~\eqref{eqn:gaussian_obs}, suppose we have a sequence of forward models
$(G_{\fidelity})_{\fidelity = 0}^{L}$, which define a
corresponding sequence of likelihoods,
$\likelyFidel(\theta) = \NN(\bfy^{*};G_{\fidelity}(\theta),\Sigma)$. Here we imagine that $G_{\fidelity+1}$
is more ``faithful'' to the data-generating process than its predecessor $G_{\fidelity}$, but
$G_{\fidelity}$ is cheaper to query; e.g., $G_{\fidelity}$ is simulated with fewer computational degrees
of freedom or involves simplified physics relative to $G_{\fidelity+1}$. To compute expectations over the more
complicated posterior distribution induced by $G_{\fidelity+1}$, we use a biasing distribution
built from evaluations of the cheaper model $G_{\fidelity}$.
For positive definite
$\Sigma$ and reasonable models $(G_{\fidelity})_{\ell=0}^{L}$,\footnote{For instance, we
might assume that our domain $\Omega$ is compact and the models are bounded on $\Omega$.}
the IS estimator in \eqref{eqn:importance_samp} will converge almost surely when using
unnormalized target density $\likelyFidelP\prior$ and unnormalized biasing density $\likelyFidel\prior$. %

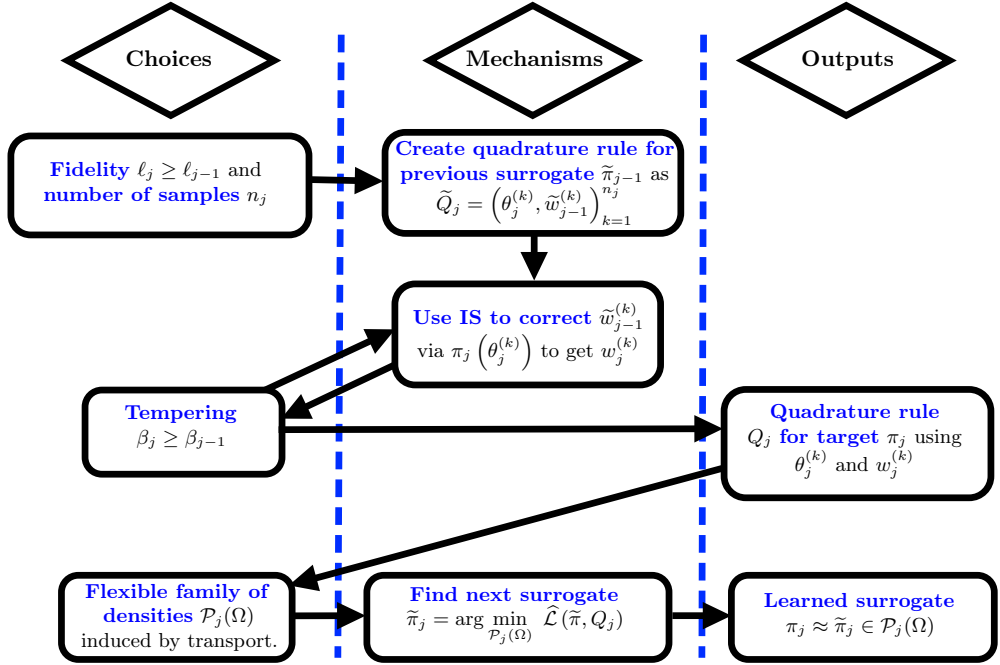
\begin{figure}[!ht]
	\centering
        \scalebox{0.85}{
    	\tikzset{every picture/.style={line width=0.75pt}} %

\begin{tikzpicture}[x=0.75pt,y=0.75pt,yscale=-1,xscale=1]

\draw [color={rgb, 255:red, 0; green, 46; blue, 255 }  ,draw opacity=1 ][line width=3]  [dash pattern={on 7.88pt off 4.5pt}]  (528.5,22) -- (527.5,392) ;
\draw [color={rgb, 255:red, 0; green, 46; blue, 255 }  ,draw opacity=1 ][line width=3]  [dash pattern={on 7.88pt off 4.5pt}]  (313.5,22) -- (311.5,392) ;
\draw  [line width=3]  (339.5,92) .. controls (339.5,85.37) and (344.87,80) .. (351.5,80) -- (503,80) .. controls (509.63,80) and (515,85.37) .. (515,92) -- (515,128) .. controls (515,134.63) and (509.63,140) .. (503,140) -- (351.5,140) .. controls (344.87,140) and (339.5,134.63) .. (339.5,128) -- cycle ;
\draw [line width=3]    (293,108) -- (332.5,108.87) ;
\draw [shift={(338.5,109)}, rotate = 181.26] [fill={rgb, 255:red, 0; green, 0; blue, 0 }  ][line width=0.08]  [draw opacity=0] (16.97,-8.15) -- (0,0) -- (16.97,8.15) -- cycle    ;
\draw [line width=3]    (345.35,218) -- (283.27,247.43) ;
\draw [shift={(277.85,250)}, rotate = 334.64] [fill={rgb, 255:red, 0; green, 0; blue, 0 }  ][line width=0.08]  [draw opacity=0] (16.97,-8.15) -- (0,0) -- (16.97,8.15) -- cycle    ;
\draw [line width=3]    (510,367) -- (539,367) ;
\draw [shift={(545,367)}, rotate = 180] [fill={rgb, 255:red, 0; green, 0; blue, 0 }  ][line width=0.08]  [draw opacity=0] (16.97,-8.15) -- (0,0) -- (16.97,8.15) -- cycle    ;
\draw [line width=3]    (428,139) -- (428,162) ;
\draw [shift={(428,168)}, rotate = 270] [fill={rgb, 255:red, 0; green, 0; blue, 0 }  ][line width=0.08]  [draw opacity=0] (16.97,-8.15) -- (0,0) -- (16.97,8.15) -- cycle    ;
\draw  [line width=3]  (427.12,4.65) -- (488.44,37.33) -- (427.12,70) -- (365.81,37.33) -- cycle ;
\draw  [line width=3]  (345.35,182) .. controls (345.35,175.37) and (350.72,170) .. (357.35,170) -- (491,170) .. controls (497.63,170) and (503,175.37) .. (503,182) -- (503,218) .. controls (503,224.63) and (497.63,230) .. (491,230) -- (357.35,230) .. controls (350.72,230) and (345.35,224.63) .. (345.35,218) -- cycle ;
\draw  [line width=3]  (117.5,93.9) .. controls (117.5,87.41) and (122.76,82.14) .. (129.25,82.14) -- (282.62,82.14) .. controls (289.12,82.14) and (294.38,87.41) .. (294.38,93.9) -- (294.38,129.16) .. controls (294.38,135.65) and (289.12,140.91) .. (282.62,140.91) -- (129.25,140.91) .. controls (122.76,140.91) and (117.5,135.65) .. (117.5,129.16) -- cycle ;
\draw  [line width=3]  (147.5,353.01) .. controls (147.5,347.49) and (151.98,343.02) .. (157.5,343.02) -- (274.38,343.02) .. controls (279.9,343.02) and (284.38,347.49) .. (284.38,353.01) -- (284.38,383) .. controls (284.38,388.52) and (279.9,393) .. (274.38,393) -- (157.5,393) .. controls (151.98,393) and (147.5,388.52) .. (147.5,383) -- cycle ;
\draw  [line width=3]  (613.16,4.65) -- (674.48,37.33) -- (613.16,70) -- (551.85,37.33) -- cycle ;
\draw  [line width=3]  (212.38,4.65) -- (273.7,37.33) -- (212.38,70) -- (151.07,37.33) -- cycle ;
\draw [line width=3]    (276.33,256.55) -- (534,256.01) ;
\draw [shift={(540,256)}, rotate = 179.88] [fill={rgb, 255:red, 0; green, 0; blue, 0 }  ][line width=0.08]  [draw opacity=0] (16.97,-8.15) -- (0,0) -- (16.97,8.15) -- cycle    ;
\draw [line width=3]    (541.5,279) -- (288.17,347.45) ;
\draw [shift={(282.38,349.01)}, rotate = 344.88] [fill={rgb, 255:red, 0; green, 0; blue, 0 }  ][line width=0.08]  [draw opacity=0] (16.97,-8.15) -- (0,0) -- (16.97,8.15) -- cycle    ;
\draw [line width=3]    (285.05,367.09) -- (322,367.01) ;
\draw [shift={(328,367)}, rotate = 179.88] [fill={rgb, 255:red, 0; green, 0; blue, 0 }  ][line width=0.08]  [draw opacity=0] (16.97,-8.15) -- (0,0) -- (16.97,8.15) -- cycle    ;
\draw  [line width=3]  (328.71,353.01) .. controls (328.71,347.49) and (333.19,343.02) .. (338.71,343.02) -- (498,343.02) .. controls (503.52,343.02) and (508,347.49) .. (508,353.01) -- (508,383) .. controls (508,388.52) and (503.52,393) .. (498,393) -- (338.71,393) .. controls (333.19,393) and (328.71,388.52) .. (328.71,383) -- cycle ;
\draw  [line width=3]  (540.67,245.53) .. controls (540.67,238.43) and (546.43,232.67) .. (553.54,232.67) -- (686.13,232.67) .. controls (693.24,232.67) and (699,238.43) .. (699,245.53) -- (699,284.13) .. controls (699,291.24) and (693.24,297) .. (686.13,297) -- (553.54,297) .. controls (546.43,297) and (540.67,291.24) .. (540.67,284.13) -- cycle ;
\draw  [line width=3]  (161.5,243.01) .. controls (161.5,237.49) and (165.98,233.02) .. (171.5,233.02) -- (268.38,233.02) .. controls (273.9,233.02) and (278.38,237.49) .. (278.38,243.01) -- (278.38,273) .. controls (278.38,278.52) and (273.9,283) .. (268.38,283) -- (171.5,283) .. controls (165.98,283) and (161.5,278.52) .. (161.5,273) -- cycle ;
\draw [line width=3]    (268.38,233.02) -- (340.06,199.54) ;
\draw [shift={(345.5,197)}, rotate = 154.97] [fill={rgb, 255:red, 0; green, 0; blue, 0 }  ][line width=0.08]  [draw opacity=0] (16.97,-8.15) -- (0,0) -- (16.97,8.15) -- cycle    ;
\draw  [line width=3]  (545,353.01) .. controls (545,347.49) and (549.48,343.02) .. (555,343.02) -- (691,343.02) .. controls (696.52,343.02) and (701,347.49) .. (701,353.01) -- (701,383) .. controls (701,388.52) and (696.52,393) .. (691,393) -- (555,393) .. controls (549.48,393) and (545,388.52) .. (545,383) -- cycle ;

\draw (184,30) node [anchor=north west][inner sep=0.75pt]  [font=\normalsize,color={rgb, 255:red, 0; green, 0; blue, 0 }  ,opacity=1 ] [align=left] {\textbf{Choices}};
\draw (386,30) node [anchor=north west][inner sep=0.75pt]  [font=\normalsize,color={rgb, 255:red, 0; green, 0; blue, 0 }  ,opacity=1 ] [align=left] {\textbf{Mechanisms}};
\draw (585,30) node [anchor=north west][inner sep=0.75pt]  [font=\normalsize,color={rgb, 255:red, 0; green, 0; blue, 0 }  ,opacity=1 ] [align=left] {\textbf{Outputs}};
\draw (203.48,110.61) node  [font=\small,color={rgb, 255:red, 0; green, 0; blue, 0 }  ,opacity=1 ] [align=left] {\begin{minipage}[lt]{121.02pt}\setlength\topsep{0pt}
\begin{center}
\textcolor{blue}{\textbf{Fidelity}} $\displaystyle \ell _{\stepIDX} \geq \ell _{\stepIDX-1}$ and\\\textcolor{blue}{\textbf{number of samples}} $\displaystyle n_{\stepIDX}$
\end{center}

\end{minipage}};
\draw (165,240.55) node [anchor=north west][inner sep=0.75pt]  [font=\small,color={rgb, 255:red, 0; green, 0; blue, 0 }  ,opacity=1 ] [align=left] {\begin{minipage}[lt]{77.74pt}\setlength\topsep{0pt}
\begin{center}
\textcolor{blue}{\textbf{Tempering}}\\$\displaystyle \beta _{\stepIDX} \geq \beta _{\stepIDX-1}$
\end{center}

\end{minipage}};
\draw (155,346.5) node [anchor=north west][inner sep=0.75pt]  [font=\small,color={rgb, 255:red, 0; green, 0; blue, 0 }  ,opacity=1 ] [align=left] {\begin{minipage}[lt]{90pt}\setlength\topsep{0pt}
\begin{center}
\textcolor{blue}{\textbf{Flexible family of}}\\\textcolor{blue}{\textbf{densities}} $\displaystyle \mathcal{P}_\stepIDX( \Omega )$\\ induced by transport.
\end{center}

\end{minipage}};
\draw (428.25,110) node  [font=\small,color={rgb, 255:red, 0; green, 0; blue, 0 }  ,opacity=1 ] [align=left] {\begin{minipage}[lt]{125pt}\setlength\topsep{0pt}
\begin{center}
\textcolor{blue}{\textbf{Create quadrature rule for}}\\\textcolor{blue}{\textbf{previous surrogate}} $\displaystyle \surrogate_{\stepIDX-1}$ as\\$\displaystyle \widetilde{Q}_{\stepIDX} =\left( \theta _{\stepIDX}^{( k)} ,\widetilde{w}_{\stepIDX-1}^{( k)}\right)_{k=1}^{n_{\stepIDX}}$
\end{center}

\end{minipage}};
\draw (352,180) node [anchor=north west][inner sep=0.75pt]  [font=\small,color={rgb, 255:red, 0; green, 0; blue, 0 }  ,opacity=1 ] [align=left] {\begin{minipage}[lt]{106.81pt}\setlength\topsep{0pt}
\begin{center}
\textcolor{blue}{\textbf{Use IS to correct}} $\displaystyle \widetilde{w}_{\stepIDX-1}^{( k)}$\\via $\displaystyle \posterStep\left( \theta _{\stepIDX}^{( k)}\right)$ to get $\displaystyle w_{\stepIDX}^{( k)}$
\end{center}

\end{minipage}};
\draw (333.85,348.29) node [anchor=north west][inner sep=0.75pt]  [font=\small,color={rgb, 255:red, 0; green, 0; blue, 0 }  ,opacity=1 ] [align=left] {\begin{minipage}[lt]{121.13pt}\setlength\topsep{0pt}
\begin{center}
\textcolor{blue}{\textbf{Find next surrogate}}\\$\displaystyle \surrogateStep =\arg\min_{\mathcal{P}_\stepIDX( \Omega )} \ \widehat{\mathcal{L}}\left(\surrogate ,Q_{\stepIDX}\right)$
\end{center}

\end{minipage}};
\draw (557,352) node [anchor=north west][inner sep=0.75pt]  [font=\small,color={rgb, 255:red, 0; green, 0; blue, 0 }  ,opacity=1 ] [align=left] {\begin{minipage}[lt]{93.55pt}\setlength\topsep{0pt}
\begin{center}
\textcolor{blue}{\textbf{Learned surrogate}}\\$\displaystyle \posterStep \approx \surrogateStep \in \mathcal{P}_\stepIDX( \Omega )$
\end{center}

\end{minipage}};
\draw (544,240) node [anchor=north west][inner sep=0.75pt]  [font=\small] [align=left] {\begin{minipage}[lt]{108.98pt}\setlength\topsep{0pt}
\begin{center}
\textcolor{blue}{\textbf{Quadrature rule}}\\$\displaystyle Q_{\stepIDX}$ \textcolor{blue}{\textbf{for target}} $\displaystyle \posterStep$ using\\$\displaystyle \theta _{\stepIDX}^{( k)}$ and $\displaystyle w_{\stepIDX}^{( k)}$
\end{center}

\end{minipage}};

\end{tikzpicture}
        }
	\caption{Illustration of a single step $\stepIDX$ of our method, proceeding from top left
	to bottom right. %
    The relationship between choosing $\temperStep$ and reweighting $\widetilde{Q}_\stepIDX$ is bidirectional: we select parameter $\temperStep$ to ensure high-quality weights $w_\stepIDX^{(k)}$.
	}
	\label{fig:flowchart}
\end{figure}%

\subsection{Core sequential methodology}
\label{sec:core_method_outline}
Now we describe each iteration of our generalized annealing scheme, integrating tempering and multi-fidelity modeling with sequential IS.
We define a sequence of integer fidelities and real-valued inverse temperatures $(\fidelity_{\stepIDX},\beta_{\stepIDX})_{j=0}^{M}\subset \mathbb{N}\times [0,1]$, for steps of the generalized annealing process $\stepIDX=0,\ldots,M$, satisfying $0=\fidelity_{0}\leq \fidelity_{1}\leq \cdots\leq \fidelity_{M}  = L$ and $0 < \temper_{0}< \temper_{1}<\cdots < \temper_{M}= 1$. Then we have a sequence of target distributions:
\begin{equation}
	\label{eqn:sequential_posteriors}
		\poster_{\stepIDX}(\theta)                                      \propto \left[\likelyStep(\theta)\right]^{\temperStep}\prior(\theta).
\end{equation}
We will approximate each element of this sequence in succession. The target distributions $(\poster_{\stepIDX})_{\stepIDX=1}^{M}$ thus induce
a corresponding sequence of biasing distributions $(\surrogateStep
)_{\stepIDX=1}^{M}$, where each $\surrogateStep$ approximates $\posterStep$ by minimizing the
$\stepIDX$-th cross-entropy loss
$\mathcal{L}_{\stepIDX}[\surrogate] := -\EE_{\posterStep}\left[\log \surrogate\right]$ over biasing distributions $\surrogate \in \mathcal{P}_{\stepIDX}(\Omega)$, %
where $\mathcal{P}_{\stepIDX}(\Omega)\subseteq \mathcal{P}(\Omega)$ is some family of distributions%
. Crucially, each such loss is estimated via importance sampling with the \textit{previous} biasing distribution, i.e.,
\begin{equation}
\label{eqn:iteratedIS}
\mathcal{L}_{\stepIDX}[\surrogate] = - \int\log\surrogate(\theta) \,\mathrm{d}\posterStep(\theta)  = -\int \log \surrogate(\theta)  \frac{\posterStep(\theta)}{\surrogate_{\stepIDX-1}(\theta)} \mathrm{d} \surrogate_{\stepIDX - 1}(\theta). %
\end{equation}
The next sections show how we discretize the loss $\mathcal{L}_\stepIDX$ via quadrature rules, i.e., sets of
points and weights. \ymmtext{\Cref{fig:flowchart} is a schematic of our overall methodology, highlighting the modular steps one must take within each iteration of the sequence.}

\section{Measure transport}
\label{sec:transport}

Now we discuss the families of
distributions, $\mathcal{P}_\stepIDX(\Omega)$, from which we select our biasing distributions
$\surrogateStep$. As noted in \Cref{sec:formulation}, we must be able to sample from these biasing
distributions and evaluate their densities. For good performance, and since ultimately we want
$\surrogate_{M}$ to be an accurate standalone approximation of $\poster$, we need each $\mathcal{P}_\stepIDX(\Omega)$ to include highly anisotropic, complicated distributions.
For these reasons---and for additional reasons involving transformations of quadrature rules,
detailed in \Cref{sec:numerical_integration}---we construct our biasing distributions via measure transport. Transport
makes the optimization in~\eqref{eqn:loss_minimization} %
tractable while allowing for particularly expressive surrogates. \ymmtext{Our approach to measure transport entails the creation of deterministic \textit{transport maps}.}

\subsection{Background on transport maps}
A transport map $S:\Omega\to\Omega$ is, at its heart, a change of variables. Conceptually, we aim to
transform a simple probability distribution $\refer$, which we are able to choose, into
our given target distribution $\poster$. This \textit{reference} distribution $\eta$, defined on
$\Omega$, is simple in the sense that we can easily sample or integrate with respect to it and that it
has a tractable density; a uniform distribution on the hypercube or a Gaussian distribution are
common choices. Then we choose an \textit{invertible} transport map $S$ to satisfy the change of
variables,%
\footnote{$U$-substitution, often introduced in elementary calculus courses, is the one-dimensional
analog of this transformation.}
\[
	\EE_{\poster}[h] = \int h(\theta)\,\poster(\mathrm{d}\theta) = \int h(S^{-1}(z) )\,\refer(\mathrm{d}
	z) = \EE_{\refer}[h\circ S^{-1}].
\]
In other words, %
if we have a random variable $\Theta\sim\poster$ and define $Z:= S(\Theta)$, then $Z$ satisfies $Z\sim
\refer$. The map $S$ is then said to \textit{push forward} $\poster$ to $\refer$, a relationship
denoted by $S_{\sharp}\poster=\refer$. Equivalently, if we have a random variable $Z\sim\refer$ and
let $\Theta := S^{-1}(Z)$,
then $\Theta\sim\poster$. We say that $\poster$ is the \textit{pullback} of $\refer$ under $S$,
denoted $S^{\sharp}\refer=\poster$. A map $S$ coupling $\poster$ and $\eta$ in this way can always be
found if $\poster$ and $\refer$ are absolutely continuous \cite{villaniOptimalTransportOld2009}. If
$S$ is also differentiable, then the change-of-variables yields a simple formula for the pullback
density,
\begin{equation}
	\label{eqn:pullback_def}\pi(\theta) = [S^{\sharp}\refer](\theta) := \refer(S(\theta ))|\nabla S(\theta
	)|,
\end{equation}
where $|\nabla S(\theta)|$ is the determinant of the Jacobian of $S$ evaluated at $\theta$.\footnote{The
inverse function theorem similarly gives
$S_{\sharp}\poster(z) = \poster(S^{-1}(z))|\nabla S(S^{-1}(z))|^{-1}$.} %
Henceforth, we will only consider transport maps $S$ that are invertible and differentiable.
Given an invertible transport $S$, we can approximate posterior expectations $\EE_{\poster}[h] = \EE_{\refer}
[h\circ S^{-1}]$ via any quadrature scheme for $\refer$. %
Because we choose $\refer$ to allow such integration rules, but in general cannot do so with $\poster$, the map $S$ makes these expectations tractable.
For more
background on measure transport, see~\cite{baptistaRepresentationLearningMonotone2023,marzoukSamplingMeasureTransport2016,villaniOptimalTransportOld2009,ramgraberFriendlyIntroductionTriangular2025}.

Now we split the problem of finding a transport map into two parts. The first question is how to
choose a suitable class of invertible functions $\FF$ within which we will search for maps. The second
question is, for a given target $\poster$, how to identify a best $S$ within this class---i.e., an $S
\in \FF$ such that $S^{\sharp} \refer$ is a good approximation of $\poster$ in the sense of minimizing
forward KL divergence. Note that $\eta$ and $\FF$ determine $\mathcal{P}(\Omega)$, i.e., $\mathcal{P}
(\Omega) = \{ S^{\sharp} \eta : S \in \FF \}$. To emphasize the present contributions, we defer discussion
of the first question to \Cref{app:map_param},
as there is ample previous literature on how to
construct useful $\FF$.
Here we focus on the second question, assuming a given fixed set of candidate transports. %

\subsection{Cross-entropy minimization with transport}
\label{sec:learning_map}
To approximate and minimize the cross-entropy loss $\mathcal{L}[\surrogate]$ over densities
$\surrogate$ induced by transport maps, we first consider the generic loss in \Cref{eqn:loss_minimization}.
We assume that $\FF$ is a finite-dimensional set, each element of which is described by parameters $\bfc
\in\RR^{p}$, $p < \infty$. The reference distribution $\refer$ is also fixed. We then write
$\surrogate(\cdot;\bfc) = S(\cdot;\bfc)^{\sharp}\refer$ for any element of $\mathcal{P}(\Omega)$. Using
\eqref{eqn:pullback_def}, minimization of the cross-entropy objective \eqref{eqn:loss_minimization}
can be written
\begin{subequations}
	\label{eqn:cross_entropy}
	\begin{align}
		\label{eqn:constrained_cross_entropy}\bfc^{*}\in \arg   & \min_{\bfc\in\RR^p}\ \mathcal{L}[\surrogate(\cdot;\,\bfc)],     \\
		\label{eqn:CEwithpullback}\mathcal{L}[\surrogate(\cdot;\,\bfc)] = -\EE_{\Theta \sim \poster}[\log \surrogate(\Theta;\bfc)] & = -\EE_{\Theta \sim \poster}[ \log\refer( S(\Theta;\bfc)) + \log|\nabla S(\Theta;\bfc)|].
	\end{align}
\end{subequations} %
Now suppose that we have a weighted-sample approximation of expectations with respect to $\poster$, given
by a quadrature rule $Q =(\theta^{(k)},w^{(k)})_{k=1}^{n}$. Then we can make the approximation,
\begin{equation}
	\label{eqn:cross_entropy_quad}\mathcal{L}[\surrogate(\cdot\,;\bfc)]\approx \widehat{\mathcal{L}}(\bfc
	;\,Q) := -\sum_{k=1}^{n}w^{(k)}\left[\log\refer\circ S(\theta^{(k)};\bfc) + \log|\nabla S(\theta^{(k)}
	;\bfc)|\right].
\end{equation}
The choice of quadrature rule $Q$ subsumes many common methods, e.g., Gaussian quadratures, quasi-Monte
Carlo schemes. If the weights are uniform and the $\theta^{(k)}$ are drawn i.i.d.\ from $\poster$,
then $Q$ is a simple Monte Carlo approximation of the expectation. In this case, loss
$\widehat{\mathcal{L}}$ can also be interpreted as a negative log-likelihood function, such that its
minimizer $\bfc^{\ast}$ is a maximum likelihood estimate of the coefficients $\bfc$ and hence of the
map $S \in \FF$ \cite{wangMinimaxDensityEstimation2022,baptistaRepresentationLearningMonotone2023,ramgraberFriendlyIntroductionTriangular2025}.

In general, once the loss is discretized as in \eqref{eqn:cross_entropy_quad}, we identify the map by
minimizing $\widehat{\mathcal{L}}( \bfc ; Q)$ with respect to $\bfc$; gradients
$\nabla_{\bfc}\, \widehat{\mathcal{L}}( \bfc ; Q)$ are easily computed and a variety of gradient-based
optimization algorithms can then be applied. Crucially, computing these gradients does not involve
evaluating the gradient of the posterior density $\poster$. Moreover, no further evaluations of the posterior density are required during the optimization iterations.

\subsection{Sequential, importance-reweighted, transport map construction}
\label{sec:numerical_integration}
Next we bring this construction into the sequential framework described by \eqref{eqn:iteratedIS}. We
replace the target $\poster$ defining the cross-entropy loss $\mathcal{L}$ above with the annealed
target distribution at step $\stepIDX$, $\posterStep$, defined in \eqref{eqn:sequential_posteriors}. Now our goal
is to approximate the intermediate loss $\mathcal{L}_{\stepIDX} [\surrogate] = -\mathbb{E}_{\posterStep}
[\log\surrogate]$.
An expectation over $\posterStep$ is generally intractable, but \eqref{eqn:iteratedIS} converts it
to an expectation over the surrogate
$\surrogate_{\stepIDX-1}{\equiv \surrogate_{\stepIDX-1}(\cdot \, ; \bfc_{\stepIDX-1}^\ast)}$, which approximates
the previous target distribution $\poster_{\stepIDX-1}$.
Recalling~\eqref{eqn:cross_entropy_quad},
we thus wish to define $Q_{j} = (\theta_{\stepIDX}^{(k)}, w_{\stepIDX}^{(k)})_{j=1}^{N_\stepIDX}$ for $\posterStep$ given a quadrature rule for $\surrogate_{\stepIDX-1}$.

Below we will show how to construct a quadrature rule
$\surrQuad_{\stepIDX-1}= (\widetilde{\theta}_{\stepIDX-1}^{(k)},\refweight_{\stepIDX-1}^{(k)})_{k=1}^{N_\stepIDX}$
for $\surrogate_{\stepIDX-1}$, taking advantage of the fact that the surrogate is specified by a transport
map. But first, suppose that we have such a rule. Then the cross-entropy loss \eqref{eqn:iteratedIS}
at step $\stepIDX$ becomes:
\begin{align}
	\label{eqn:reweightedCEloss}\mathcal{L}_{\stepIDX}[\surrogate( \cdot; \bfc)] & = -\mathbb{E}_{\Theta \sim\posterStep}[\log \surrogate(\Theta ; \bfc)] = -\mathbb{E}_{\Theta \sim \surrogate_{\stepIDX-1}}\left [ \log \surrogate(\Theta; \bfc) \frac{\posterStep(\Theta)}{\surrogate_{\stepIDX-1}(\Theta)}\right ] \\[-0.5em]
	     & \approx \widehat{\mathcal{L}}_{j}(\bfc; Q_{\stepIDX}) := -\sum_{k=1}^{N_\stepIDX}w_{\stepIDX}^{(k)}\log\surrogate (\theta_{\stepIDX}^{(k)};\bfc), \nonumber \\
	     & \qquad \qquad \quad\ \ = - \sum_{k=1}^{N_j}w_{j}^{(k)}\left ( \log\refer(S(\theta^{(k)}_{j}; c)) + \log|\nabla S(\theta^{(k)}_{j}; c)| \right ), \label{subeqn:reweight_pullback}\\%\ \  . \nonumber
         \label{eqn:reweigh}  \text{where }\ \ \theta_{\stepIDX}^{(k)}&= \widetilde{\theta}_{\stepIDX-1}^{(k)}, \ \
    \nu_{\stepIDX}^{(k)}:=\refweight_{\stepIDX-1}^{(k)}\ \frac{\posterStep}{\surrogate_{\stepIDX - 1}}(\theta_{\stepIDX}
	^{(k)}),\ \
    w_{\stepIDX}^{(k)} := \frac{1}{\sum_{k=1}^{n_j} \unnormw_\stepIDX^{(k)}}\unnormw_\stepIDX^{(k)}.
\end{align}
\Cref{subeqn:reweight_pullback} comes from inserting the pullback $S^{\sharp}\refer$ from \eqref{eqn:CEwithpullback}
into $\widehat{\mathcal{L}}_\stepIDX$.
Minimizing $\widehat{\mathcal{L}}_{j}(\bfc; Q_{\stepIDX})$ over $\bfc \in \mathbb{R}^{p}$ yields the
coefficients $\bfc_{\stepIDX}^{\ast}$ of the surrogate
$\surrogateStep(\cdot ; \bfc_{\stepIDX}^{\ast})$.
And the weights $(w_{\stepIDX}^{(k)})_{k=1}^{N_\stepIDX}$ capture the fact that we evaluate the loss
$\mathcal{L}_{\stepIDX}$ by \textit{reweighing} the quadrature rule $\surrQuad_{\stepIDX-1}$ to
produce $Q_{\stepIDX}$.

We now turn to the construction of $\surrQuad_{\stepIDX-1}$, the quadrature rule for the preceding surrogate
distribution $\surrogate_{\stepIDX-1}$.
Developing deterministic quadrature schemes, e.g., Gaussian quadratures, for general multivariate distributions is difficult~\cite{gautschi1994algorithm, ma1996generalized, gander2001stable}.
Crucially, however, our surrogate distribution is explicitly constructed as the pullback of a
standard/tractable reference distribution, i.e., $\surrogate_{\stepIDX-1}= S_{\stepIDX - 1}^{\sharp}\refer$.
Let $(z^{(k)}, w_\refer^{(k)})_{k=1}^{N_\stepIDX}$
be an $N_{\stepIDX}$-point
quadrature rule for $\refer$. We can then use the invertible map $S_{\stepIDX - 1}$ to define a
quadrature rule for $\surrogate_{\stepIDX-1}$ as
$\surrQuad_{\stepIDX-1}= (\widetilde{\theta}_{\stepIDX-1}^{(k)},\refweight_{\stepIDX-1}^{(k)})_{k=1}^{N_{\stepIDX}}$
with
\begin{equation}
	\label{eqn:transport_quad}\widetilde{\theta}_{\stepIDX-1}^{(k)}= S_{\stepIDX - 1}^{-1}(z^{(k)}), \
	\ \refweight_{\stepIDX-1}^{(k)}= w_\refer^{(k)}.
\end{equation}
In other words, the abscissae of the reference quadrature rule are transported and its weights are preserved~\cite{ernstLearningIntegrate2025,stroudApproximateCalculationMultiple1971}.
These abscissae and weights directly enter the expression \eqref{eqn:reweigh} above, where they are reweighted
to account for the mismatch between $\surrogate_{\stepIDX-1}$ and $\posterStep$. Note that if the
reference quadrature is unweighted independent Monte Carlo, $\surrQuad_{\stepIDX-1}$ amounts to i.i.d.\ sampling
from $\surrogate_{\stepIDX-1}$. But if the reference quadrature is chosen differently, as we will
demonstrate in our numerical examples, the pullback quadrature can yield a more sample-efficient approximation
of the loss $\mathcal{L}_{\stepIDX}$.

A straightforward approach to creating the sequence $(\surrogateStep)_{\stepIDX=0}^{M}$, as outlined
in \Cref{fig:flowchart}, is to use a new reweighted pullback quadrature rule $Q_\stepIDX$ to define the loss
$\widehat{\mathcal{L}}_{j}$ at each step $\stepIDX=0,\ldots,M$, with the convention $\surrogate_{-1}\equiv
\prior$. %
The parameters $\bfc^{*}_{\stepIDX}$ of the next biasing distribution $\surrogateStep$ are found by minimizing
$\widehat{\mathcal{L}}_{j}(\cdot\,;\,Q_{\stepIDX})$, and any minimizer $\bfc^{*}_{\stepIDX}$ of this
function is invariant to the normalization of $\posterStep$. Moreover, we can evaluate $\posterStep$
at all $N_{\stepIDX}$ quadrature points in parallel; compare this with MCMC, where parallel evaluation
is often impossible within a single chain. %
At a high level, our iterative approach connects successive steps, $\stepIDX$ and $\stepIDX-1$, with
importance weighting and connects each step to a common reference distribution through transport.

\section{Algorithm and remarks}
\label{sec:algorithms} As a final step towards our full algorithm, we discuss a more complex method of
building the quadrature rule $Q_{\stepIDX}$ at each step. While we can use~\eqref{eqn:reweigh} and~\eqref{eqn:transport_quad} to calculate $\widehat{\mathcal{L}}_{\stepIDX}(\cdot \, ; Q_{\stepIDX})$ (and its gradients)
at each step, this approach na\"{i}vely discards presumably expensive evaluations of the
likelihood $\likely$. Below, we will show that these evaluations can be actually be re-used across
tempering steps, i.e., when reweighting quadrature rules via~\eqref{eqn:reweigh}. By reusing these evaluations, we improve the quality
of our transport map approximations, which can otherwise be limited by small sample sizes. We then explore choices of the reference  distriubtion $\eta$, list our complete algorithms, and contextualize our work on the frontier of Bayesian inference. %

\subsection{Multiple tempered importance-weighted quadrature}
\label{sec:multiple_tempering}%
To illustrate our re-use scheme, we first consider two successive steps of the iterative algorithm described in \Cref{sec:numerical_integration}. Let these steps have likelihoods of the same fidelity, i.e., $\fidelity_{j} = \fidelity_{j+1}= \fidelity$, but different tempering parameters $\beta_{j} < \beta_{j+1}$.
The quadrature rule $Q_{j}$ used to evaluate $\widehat{\mathcal{L}}_{j}(\cdot \, ; Q_{j})$ begins with a quadrature rule $\surrQuad_{\stepIDX-1}= (\widetilde{\theta}_{\stepIDX-1}^{(k)},\refweight_{\stepIDX-1}^{(k)})_{k=1}^{n_\stepIDX}$ that is then reweighted according to \eqref{eqn:reweigh}. Writing this process more explicitly, we first compute the \textit{unnormalized weights} $\unnormw_{j}^{(k)}$, and then normalizing
them to form $Q_{j}$:%
\begin{gather}\label{eqn:jweights}
	\unnormw_{\stepIDX}^{(k)}:= \refweight_{\stepIDX-1}^{(k)}\frac{\left( u_{j}^{(k)}\right)^{\beta_\stepIDX}}{v_{j}^{(k)}}, \ \text{where}\ u_{j}^{(k)}:= \likely_{\fidelity}(\theta_{\stepIDX}^{(k)}), \ \ v_{j}^{(k)}:=\frac{\surrogate_{\stepIDX - 1}(\theta_{\stepIDX}^{(k)})}{\prior (\theta_{\stepIDX}^{(k)})}, \ \ \theta_{\stepIDX}^{(k)}= \widetilde{\theta}_{\stepIDX-1}^{(k)},\\
	Q_{j} = \left ( \theta_{\stepIDX}^{(k)}, \frac{1}{W_{\stepIDX}}\unnormw_{\stepIDX}^{(k)}\right )_{k=1}^{n_\stepIDX}, \ \ \text{with}\ W_{j} = \sum_{k=1}^{n_\stepIDX}\unnormw_{\stepIDX}^{(k)}.\nonumber
\end{gather}
Note that this process entails $n_{j}$ evaluations of the likelihood $\likely_{\fidelity}$.
Next, we minimize $\widehat{\mathcal{L}}_{j}(\cdot \, ; Q_{j})$ to find a new surrogate $\surrogate_{j}$ described by a transport map $S_{j}$. Via this transport, we construct a new quadrature rule $\surrQuad_{\stepIDX}= (\widetilde{\theta}_{\stepIDX}^{(k)},\refweight_{\stepIDX}^{(k)})_{k=1}^{n_{\stepIDX+1}}$. To obtain $Q_{j+1}$ from $\surrQuad_{j}$, we compute new unnormalized weights then normalize them:%
\begin{gather}\label{eqn:jp1weights}
\unnormw_{\stepIDX+1}^{(k)}:= \refweight_{\stepIDX}^{(k)}\frac{\left( u_{j+1}^{(k)}\right)^{\beta_{j+1}}}{v_{j+1}^{(k)}}, \ \text{where}\ u_{j+1}^{(k)}:= \likely_{\fidelity}(\theta_{\stepIDX+1}^{(k)}), \ \  v_{j+1}^{(k)}:= \frac{\surrogate_{\stepIDX}(\theta_{\stepIDX+1}^{(k)})}{\prior (\theta_{\stepIDX+1}^{(k)})}, \ \  \theta_{\stepIDX+1}^{(k)}= \widetilde{\theta}_{\stepIDX}^{(k)},\\
	Q_{j+1}= \left( \theta_{\stepIDX+1}^{(k)}, \frac{1}{W_{\stepIDX+1}}\unnormw_{\stepIDX+1}^{(k)}\right)_{k=1}^{n_{j+1}}, \ \ \text{with}\ W_{j+1}= \sum_{k=1}^{n_{j+1}}\unnormw_{\stepIDX+1}^{(k)}.\nonumber
\end{gather}
As evident from \eqref{eqn:jp1weights}, this process evaluates $\likely_{\fidelity}$ at $n_{j+1}$ new points.

Remarkably, the weights of both $Q_{\stepIDX}$ and $Q_{\stepIDX+1}$ use evaluations of the \emph{same} likelihood function $\likely_{\fidelity}$. Rather than applying $Q_{\stepIDX+1}$ on its own, we can instead use \textit{multiple importance sampling}~\cite{veach1998robust,owenMCBook2013}, a technique that simultaneously exploits multiple biasing distributions, to combine $Q_{\stepIDX}$ with $Q_{\stepIDX+1}$. Our combined rule $Q^{\MIS}_{\stepIDX+1}$, with $N_{\stepIDX+1}:= n_{\stepIDX}+ n_{\stepIDX+1}$ points, uses integral discretizations $Q_{\stepIDX}$ (with biasing distribution $\surrogate_{\stepIDX-1}$) and $Q_{\stepIDX+1}$ (with biasing distribution $\surrogate_{\stepIDX}$) to approximate an expectation over $\poster_{\stepIDX+1}$. In other words, we obtain a quadrature rule of size $n_{\stepIDX}+n_{\stepIDX+1}$ for step $\stepIDX+1$, but only evaluate the likelihood an additional $n_{\stepIDX+1}$ times. For an arbitrary function $h$, a multiple importance sampling estimator of its expectation over $\poster_{\stepIDX+1}$ is given by
\begin{equation}\label{eqn:example_MIS}
\EE_{\poster_{j+1}}[h]\approx%
\frac{1}{W_{\stepIDX}^{\MIS}}\sum_{k=1}^{n_{\stepIDX}}\alpha_{\stepIDX}(\theta_{\stepIDX}^{(k)})  \unnormw_{j}^{(k)}(u_{\stepIDX}^{(k)})^{\beta_{j+1} - \beta_\stepIDX}h(\theta_{\stepIDX}^{(k)}) +%
\frac{1}{W_{\stepIDX+1}^{\MIS}}\sum_{k=1}^{n_{\stepIDX+1}}\alpha_{\stepIDX+1}(\theta_{\stepIDX+1}^{(k)}) \unnormw_{\stepIDX+1}^{(k)} h(\theta_{\stepIDX+1}^{(k)}),
\end{equation}
where $W_{\stepIDX}^{\MIS}$ and $W_{\stepIDX+1}^{\MIS}$ are chosen to ensure a normalized quadrature rule\footnote{In particular, we set $W_{\stepIDX}^{\MIS} = \sum_{k=1}^{n_\stepIDX}\alpha_{\stepIDX}(\theta_{\stepIDX}^{(k)})\nu_{\stepIDX}^{(k)}(u_{\stepIDX})^{\beta_{\stepIDX+1}-\beta_{\stepIDX}}$ and $W_{\stepIDX+1}^{\MIS} = \sum_{k=1}^{n_{\stepIDX+1}}\alpha_{\stepIDX+1}(\theta_{\stepIDX+1}^{(k)})\nu_{\stepIDX+1}^{(k)}.$} and $\alpha_{\stepIDX}(\theta) + \alpha_{\stepIDX+1}(\theta) \equiv 1$ for any fixed input $\theta\in\RR^{d}$, as in~\cite{owenMCBook2013}. The first sum on the right-hand side of~\eqref{eqn:example_MIS} reflects that %
\[(u_{\stepIDX}^{(k)})^{\beta_{j+1} - \beta_\stepIDX}= \frac{\left[\likely_{\fidelity}(\theta^{(k)}_{\stepIDX})\right]^{\temper_{\stepIDX+1}}}{\left[\likely_{\fidelity}(\theta^{(k)}_{\stepIDX})\right]^{\temper_{\stepIDX}}}= \frac{\left[\likely_{\fidelity}(\theta^{(k)}_{\stepIDX})\right]^{\temper_{\stepIDX+1}}\prior(\theta^{(k)}_{\stepIDX})}{\left[\likely_{\fidelity}(\theta^{(k)}_{\stepIDX})\right]^{\temper_{\stepIDX}}\prior(\theta^{(k)}_\stepIDX)}\propto \frac{\poster_{j+1}(\theta_{j}^{(k)})}{\poster_{j}(\theta_{j}^{(k)})},\]
where the constant of proportionality comes from the choice of tempering parameters $\temper_{\stepIDX+1}$
and $\temperStep$. Note that if $(\theta_{\stepIDX}^{(k)})_{k}$ and
$(\theta_{\stepIDX+1}^{(k)})_{k}$ were truly independent
Monte Carlo samples (with $\theta_{\stepIDX}^{(k)}\sim \surrogate_{j-1}$, $\theta_{\stepIDX+1}^{(k)}
\sim \surrogate_{\stepIDX}$) and if $\surrogate_{\stepIDX}$ was itself independent of the samples
$(\theta_{\stepIDX}^{(k)})_k$, then
it is easy to show that \eqref{eqn:example_MIS} is a consistent estimator of
$\EE_{\poster_{\stepIDX+1}}[h]$~\cite{veach1998robust}. %
In practice, however, we have created $S_{\stepIDX}$ by minimizing $\widehat{\mathcal{L}}(\cdot ; Q_{\stepIDX}
)$, and thus $Q_{\stepIDX+1}$ is dependent on $Q_{\stepIDX}$ via $\surrogate_{\stepIDX}$.

Following this two-step illustration, we now consider the general case: reusing likelihood
evaluations, and hence quadrature rules $Q_{\stepIDX}$, across \textit{more} than two successive tempering steps.
These quadrature rules interact through the partition of unity $(\alpha_{i})_{i=0}^{\stepIDX}$.
Note that the solution of the optimization problem~\eqref{eqn:cross_entropy} is independent of the sum
across all weights within $Q_{\stepIDX}^{\MIS}$,
i.e., the loss function~\eqref{subeqn:reweight_pullback} is invariant to the quantity $\sum_{i=0}^{j}W
_{i}^{\MIS}$. %
Yet variance in the normalization $W_{i}$ of each individual quadrature rule $Q_{i}$, for
$i\leq \stepIDX$, could contribute to variance in the minimizer of
$\widehat{\mathcal{L}}(\cdot;\,Q^{\MIS}_{\stepIDX})$.
We use the power heuristic
$\alpha_{i}(\theta)\propto \left(n_{i}\surrogate_{i - 1}(\theta )\right)^{\gamma}$ to construct our partition of unity, where proportionality ensures a unit sum
across $i=0,\dots,j$ for  {any fixed value of} $\theta$. Following~\cite{veach1998robust}, we set the hyperparameter to $\gamma=2$, which reweighs each surrogate according to its contribution to the variance of all the importance weights. This choice balances exploration of the domain with exploitation of areas that the current surrogates do not already fit well. %
We must ensure that evaluations $(u_{i}^{(k)})_{k=1}^{n_i}$, $i \leq j$ are from one likelihood
model $\likely_{\fidelity}$; once we transition between likelihood fidelities, we restart the scheme.
Our full \textit{multiple tempered importance-weighted quadrature} approach is in \Cref{alg:multiple_IS}.

\begin{algorithm}
	\caption{Multiple tempered importance-weighted quadrature}
	\label{alg:multiple_IS}
	\begin{algorithmic}
		\REQUIRE Parameter $\temper$, memos
		$\left(\surrogate_{i},D_{i}\right)_{i=\stepIDX_0}^{\stepIDX-1}$ with
		$D_{i}=\left(\likely(\widetilde{\theta}_{i}^{(k)}), \widetilde{\theta}_{i}^{(k)}, \refweight_{i}^{(k)}\right)_{k=1}^{n_{i+1}}$

		\ENSURE Quadrature rule $Q^{\MIS}_{\stepIDX}$ for pdf $\pi_{\stepIDX}\propto\likelyStep\prior$,
		where $|Q^{\MIS}_{\stepIDX}| = N_{\stepIDX}:= \sum_{i=\stepIDX_0}^{\stepIDX-1}n_{i+1}$

		\FOR{$i=\stepIDX_{0},\ldots,\stepIDX-1$}

		\STATE Set $u_{i}^{(k)}=\likely(\widetilde{\theta}_{i}^{(k)})$ and $v_{i}^{(k)}= \frac{\surrogate_{i}}{\prior}
		(\widetilde{\theta}_{i}^{(k)})$ as the true and approximate likelihood weights, resp.

		\STATE Define $\widetilde{\alpha}_{i}(\theta) = (n_{i+1}\surrogate_{i}(\theta))^{\gamma}$ to weigh surrogate $\surrogate_i$ in MIS.%

		\STATE Define temporary weight $\widetilde{\unnormw}_{i}^{(k)}= \widetilde{\alpha}_{i}(\widetilde{\theta}^{(k)}_{i})\refweight_{i}^{(k)}\frac{[u_{i}^{(k)}]^{\beta}}{v_{i}^{(k)}}$ for abscissa $\widetilde{\theta}_i^{(k)}$.

		\STATE Store point and weight $(\widetilde{\theta}_{i}^{(k)}, \unnormw_{i}^{(k)})_{k=1}^{n_{i+1}}$ and set
		$W^{\MIS}_{i}= \sum_{k=1}^{n_{i+1}}\widetilde{\unnormw}_{i}^{(k)}$ as the normalization for $\surrogate_i$.

		\ENDFOR
		\STATE For $i=j_0,\ldots,j-1$, $k=1,\ldots,n_i$, set $\unnormw_i^{(k)} = \widetilde{\unnormw}_i^{(k)}\left(\sum_{i^\prime=j_0}^{j-1}\widetilde{\alpha}_{i^\prime}(\theta_i^{(k)})\right)^{-1}$ as MIS-normalized weight.
		\STATE Create final quadrature rule
		$Q^{\MIS}_{\stepIDX}= \bigcup_{i=\stepIDX_0}^{\stepIDX-1}\left(\theta_{i}^{(k)}, W^{-1}\ \unnormw_{i}^{(k)}\right)_{k=1}^{n_{i+1}}$, where $W = \sum_{{i}=j_0}^{\stepIDX-1}W^{\MIS}_{i}$
	\end{algorithmic}
\end{algorithm}

\subsection{Choice of reference distribution}
\label{sec:reference} %
So far, we have not specified the choice of reference distribution $\eta$ in our transport scheme, only requiring it to be a normalized density that we can sample from. Moreover, nothing in the algorithms above precludes the choice of reference from changing at each step; thus we may more generally write the reference at step $\stepIDX$ as $\refer_{\stepIDX}$.
Suppose, for the sake of illustration, that we could set $\refer_{j}= \poster_{j}$ (ignoring the fact that we cannot sample from this distribution and do not know its normalizing constant). Then the trivial map $S_{j}=\mathrm{Id}$ would perfectly capture the target at step $j$. Of course this choice is impractical, but it suggests that a good choice of reference can make the minimization of $\widehat{\mathcal{L}}_{j}(\cdot ; Q_{j})$ easier and reduce the difficulty of adequately approximating $S_{j}$~\cite{ramgraberFriendlyIntroductionTriangular2025}.
To this end, we propose that $\refer_{\stepIDX}$ is either the previous step's approximate posterior distribution $\surrogate_{\stepIDX - 1}$ or the prior distribution $\prior$. The second choice is straightforward, but the first bears further discussion.

Suppose that at step $j-1$, we have found a transport map $S_{\stepIDX-1}^{\ast}(\cdot) = S_{\stepIDX-1}(\, \cdot \, ; c^{\ast})$ achieving $( S_{\stepIDX - 1}^{\ast})^{\sharp}\refer_{\stepIDX - 1}= \surrogate_{\stepIDX - 1}\approx \poster_{\stepIDX - 1}$. Intuitively, if $\poster_{\stepIDX-1}$ and $\poster_{\stepIDX}$ are also close, then there should exist a map $S^{0}_{\stepIDX}$ achieving $\pi_{\stepIDX}= (S^{0}_{\stepIDX})^{\sharp}\surrogate_{\stepIDX-1}$ that is close to the identity (cf.\ also \cite[Lemma 4.5]{wangMinimaxDensityEstimation2022}). By setting $\eta_{\stepIDX}= \surrogate_{j-1}$ and seeking a surrogate $\surrogate_{\stepIDX}$ parameterized as \[\surrogate_{\stepIDX}= S_{\stepIDX}^{\sharp}\surrogate_{\stepIDX-1}= (S_{\stepIDX-1}^{\ast}\circ S_{\stepIDX})^{\sharp}\eta_{\stepIDX-1},\] we effectively create a composition of transport maps. We find the new map, $S_{\stepIDX}$, by minimizing $\widehat{\mathcal{L}}_{\stepIDX}(\,\cdot\,;\,Q^{\MIS}_{\stepIDX})$ as described in \Cref{sec:formulation,sec:transport}%
; at this step, the previous map $S_{\stepIDX-1}^{\ast}$ is already fixed.

We emphasize that choosing a reference distribution $\refer_{\stepIDX}$ only affects the form of $\surrogate_{\stepIDX}$ and does not change how reweighting is performed to arrive at $Q^{\MIS}_{\stepIDX}$. In other words, we design $Q^{\MIS}_{\stepIDX}$ to approximate expectations with respect to $\poster_{\stepIDX}$, independently of the reference density $\refer_{\stepIDX}$. To understand the difference between the choice of reference $\refer_{\stepIDX}$ and the quadrature rule $Q^{\MIS}_{\stepIDX}$, recall the optimization problem formulation in~\eqref{eqn:constrained_cross_entropy}. Choosing the reference changes the space of densities $\mathcal{P}_\stepIDX(\Omega)$ over which we search, whereas quadrature rule $Q^{\MIS}_{\stepIDX}$ determines the loss $\widehat{\mathcal{L}}(\cdot,Q^{\MIS}_{\stepIDX})$ that we wish to minimize.

\subsection{Complete algorithm}
\label{sec:mfinf_alg}
\begin{algorithm}
	\caption{Adaptive multi-fidelity multiple tempered importance-weighted quadrature}
	\label{alg:core_loop}
	\begin{algorithmic}
		\REQUIRE Prior distribution $\prior$, likelihood functions
		$(\likely_{\fidelity})_{\fidelity=0}^{L}$ of fidelity $\fidelity$, annealing thresholds
		$(t_{\fidelity})_{\fidelity=0}^{L}$ such that
		$0<t_{0}<t_{1}<\cdots<t_{\fidelity}<\cdots<t_{L}=1$

		\ENSURE Quadrature rule $Q^{\MIS}_{M}$ for $\pi_{M}$, transport map $S_{M}$, and reference density $\refer_{M}$ such
		that $S_{M}^{\sharp}\refer_{M}\approx \poster_{M}\propto \likely_{L}\prior$

		\STATE Initialize step counter $\stepIDX=0$, tempering parameter $\temper_{-1}=0$, and biasing
		distribution $\surrogate_{-1}\equiv\prior$

		\FOR{$\fidelity=0,\ldots,L$}

		\STATE Initialize $\stepIDX_{0}= \stepIDX$ and set
		$\refer_{\stepIDX}= \surrogate_{\stepIDX - 1}$

		\REPEAT

		\STATE Record $\fidelity_{\stepIDX}:=\fidelity$%
		\STATE \textbf{Choose} $n_{\stepIDX}$ and create quadrature rule
		$\surrQuad_{\stepIDX-1}= (\widetilde{\theta}_{\stepIDX}^{(k)},\refweight_{\stepIDX-1}^{(k)})_{k=1}
		^{n_{\stepIDX}}$
		for integration over $\surrogate_{\stepIDX - 1}$%

		\STATE Evaluate $u_{\stepIDX}^{(k)}\propto \likelyStep(\widetilde{\theta}_{\stepIDX}^{(k)})$ and
		cache
		$D_{\stepIDX}= (u_{\stepIDX}^{(k)},\widetilde{\theta}_{\stepIDX}^{(k)},\refweight_{\stepIDX-1}^{(k)}
		)_{k=1}^{n_{\stepIDX}}$%

		\STATE \textbf{Choose} $\temperStep$ such that $t_\fidelity\geq\temperStep\geq \temper_{\stepIDX-1}$%

		\STATE Create quadrature rule
		$Q^{\MIS}_{\stepIDX}= (\theta_{\stepIDX}^{(k)}, w_{\stepIDX}^{(k)})_{k=1}^{N_{\stepIDX}}$ from
		$(\surrogate_{i-1},D_{i})_{i=\stepIDX_0}^{\stepIDX}$ and $\temperStep$ via \textbf{\Cref{alg:multiple_IS}}%

		\STATE Using information on $Q^{\MIS}_{\stepIDX}$ and $\refer_{\stepIDX}$, \textbf{choose}
		regularizer $\lambda_{\stepIDX}$ and function class $\FF_{\stepIDX}$ for map
		$S_{\stepIDX}(\cdot\,;\bfc)$%

		\STATE Find parameters
		$\bfc^{*}\in \arg\min\limits_{\bfc}\widehat{\mathcal{L}}(\bfc; Q^{\MIS}_{\stepIDX}) + \lambda_{\stepIDX}
		\|\bfc\|^{2}$%
		\STATE Set $\surrogate_{\stepIDX}= S_{\stepIDX}(\cdot;\bfc^{*})^{\sharp}\refer_{\stepIDX}$,
		$\refer_{j+1}=\prior$, and increment $j$%

		\UNTIL{$\temperStep < t_{\fidelity}$}

		\ENDFOR

		\STATE Record maximum step $M = \stepIDX$
	\end{algorithmic}
\end{algorithm}
\Cref{alg:core_loop} formalizes the multi-fidelity inference method illustrated in \Cref{fig:flowchart} and detailed in Sections~\ref{sec:formulation}--\ref{sec:reference}.
The algorithm is fully described by $\gamma$, $(t_{\fidelity})_{\fidelity=0}^{L}$, $(\temperStep)_{j=0}^{M}$, $(n_{\stepIDX})_{j=0}^{M}$, $(\lambda_{\stepIDX})_{j=0}^{M}$, and $(\FF_{\stepIDX})_{j=0}^{M}$. Several of these hyperparameters are chosen in-the-loop, as emphasized by boldface type in \cref{alg:core_loop}. Our choices for setting these parameters are as follows.

As $\fidelity_{\stepIDX}$ (the likelihood fidelity level at step $j$) increases, the number of likelihood evaluations $n_{\stepIDX}$ should be chosen according to the expense of the likelihood $\likelyStep$, as suggested earlier.
The threshold sequence $(t_{\fidelity})_{\fidelity=0}^{L}$ defines the range of inverse temperature parameters $\beta_\stepIDX$ than can be used for a given fidelity; we enforce $t_{\fidelity-1} \leq \beta_\stepIDX\leq  t_\fidelity$
for steps $\stepIDX$ employing likelihood fidelity $\fidelity$, so that higher-fidelity models are deployed as we get closer to the final distribution, i.e., as $\beta_j \to 1$.
The choice of transport map function class $\FF_{\stepIDX}$ is intricately tied to the choice of quadrature rule and reference distribution. On the one hand, setting the reference to be the previous surrogate/pullback distribution, $\refer_{\stepIDX}= \surrogate_{\stepIDX-1}$, may give an easier approximation problem as described in~\Cref{sec:reference}; on the other hand, the resulting composition of transport maps can overfit to the sparse data in $Q^{\MIS}_{j}$. Such overfitting may motivate the use of more advanced {adaptive} transport map parameterizations in the future~\cite{baptistaRepresentationLearningMonotone2023}, but here we use the simple rule for choosing $\refer_{\stepIDX}$ in \Cref{alg:core_loop}, which is demonstrated to be effective in \Cref{sec:examples}: Every time a new fidelity is chosen, we set the reference to be the previous surrogate, $\refer_{\stepIDX} = \surrogate_{\stepIDX-1}$; then, after the first surrogate is found for a given fidelity, we set $\refer_{\stepIDX} = \prior$.

To choose $\beta_\stepIDX$ within its allowable range at each step and to adjust the complexity/size of $\FF_{\stepIDX}$, we compute the relative effective sample size (rESS) of the current quadrature. For any quadrature rule $Q$ with weights $\mathbf{w} = (w^{(1)},\ldots, w^{(N)})$, we define $\rESS[Q] = (N \|\mathbf{w}\|_{2}^{2})^{-1}$ %
as an a priori indicator
of error. When the weights are positive and sum to one, the rESS takes a value in
$[N^{-1},1]$, where a rESS of unity is desirable but does not guarantee high-quality integration.
If the rESS is small, many of the weights of $Q$ are near zero; if the rESS is large, the samples
are weighted nearly uniformly and thus $Q$ resembles a (quasi-)Monte Carlo quadrature. When the set of
transport maps $\FF_{\stepIDX}$ admits high-frequency functions, we avoid overfitting by using $\rESS[Q^{\MIS}_{\stepIDX}]$ to guide the choices of $\temperStep$ and $\FF_{\stepIDX}$. In general, we tune $\FF_{\stepIDX}$ to be simpler for low values of rESS.
The tempering parameters $(\beta_j)_j$ are chosen via the systematic scheme detailed in \Cref{app:temp_sched}. At a high level, this approach uses a simple (e.g., low-degree) class of functions $\FF_{\stepIDX}$ and takes only small increments of the tempering parameter $\temperStep$ (i.e., small $\temper_\stepIDX - \temper_{\stepIDX-1}$) when the rESS diagnostic is small. Conversely, we allow for larger function classes $\FF_{\stepIDX}$ and larger increases $\temper_\stepIDX - \temper_{\stepIDX-1}$ when the rESS is larger.
The $L^2$ regularization hyperparameters $(\lambda_{\stepIDX})_{j=0}^{M}$ further regularize the
transport map parameters; we find it necessary to include a regularization term for stability, but our results are insensitive to the choice of small regularization.%

\subsection{Algorithmic insight}
\label{sec:remarks}

Though we already discuss related work in \Cref{sec:relatedwork}, we can now comment more precisely on certain aspects of our algorithm.
First, recall the comparison with~\cite{farcasMultilevelAdaptiveSparse2020,liAdaptiveConstructionSurrogates2014}, which also consider the problem of highly concentrated posterior distributions. As noted earlier, a key distinction from these earlier works is the output of our algorithm: we create an expressive transport-based surrogate $\surrogate_{M}$ (which provides both a normalized density estimate and a fast i.i.d.\ sampler) and an approximate quadrature rule $Q_{M}$ for the distribution $\poster_{M}$. In contrast, \cite{liAdaptiveConstructionSurrogates2014} creates a \textit{Gaussian} approximation of $\pi$ and a polynomial approximation of the forward model; sampling from the posterior approximation induced by this forward model surrogate is then left to MCMC. \cite{farcasMultilevelAdaptiveSparse2020} uses multi-level sparse grids to create both an interpolation of the log-likelihood and a quadrature rule over an adapted Gaussian measure, but this approach is again limited by its Gaussian assumption and cannot handle strongly multi-modal targets.

More broadly, in \cref{alg:core_loop} we have specified a method with many building blocks. These building blocks are largely modular: to demonstrate our algorithm, we implement particular choices for each element, but these choices are not prescriptive or totemic.
Rather, we see our core contribution as the framework (generalized annealing, variational construction of transport, multiple tempered importance-weighted quadrature) within which all these elements interact.
One such building block is the choice of quadrature family and transport class. We find better results with non-deterministic quadrature rules, specifically randomized QMC, rather than deterministic rules. Our transports are parameterized by monotone polynomials, as detailed in \Cref{app:map_param}. Several recent papers \cite{ernstLearningIntegrate2025,jakemanPolynomialChaosExpansions2019,klebanovTransportingHigherOrderQuadrature2023,liuTransportQuasiMonteCarlo2024} have specifically studied the construction of quadrature via transport, though in contrast with some of these efforts, we use quadrature to identify transport maps (by discretizing a variational objective) \textit{and} we use transport to create new quadrature rules, rather than doing only the latter. \cite{liuTransportQuasiMonteCarlo2024} investigates the construction of transport maps specifically tailored to randomized QMC, by enforcing certain smoothness and growth rate conditions.

Another building block is the scheme for choosing a tempering schedule $(\temperStep)_{j=0}^{M}$. This is a widely studied problem, with many common heuristics and even some theory, e.g., `optimal' schedules for a fixed integrand~\cite{chopinConnectionTemperingEntropic2024,syedOptimisedAnnealedSequential2024}.
Here we have taken a practical approach, modifying the schemes in~\cite{liuAdaptiveAnnealedImportance2014,wangMitigatingModeCollapse2025}%
\footnote{These works independently proposed virtually identical methods.}
as described in \cref{app:temp_sched}. But many other tempering schedules could be incorporated in our framework.
Also, while there is ample previous work on multi-fidelity methods in importance sampling~\cite{peherstorferMultifidelityImportanceSampling2016,salehTemperedMultifidelityImportance2024,peherstorferSurveyMultifidelityMethods2018}, the use of multi-fidelity models with tempering is less widespread~\cite{salehTemperedMultifidelityImportance2024}. Very recent work~\cite{cerou2025adaptive} lends theoretical support to our heuristic of picking more complex models at higher inverse temperatures.

It is interesting to contrast the computational patterns of minimizing the cross-entropy (forward KL) loss $\mathcal{L}$, as we do here, with those of reverse KL minimization, often used in variational inference. We incur one forward model or likelihood evaluation for each quadrature point used to approximate $\mathcal{L}$; optimization iterations then involve no new model evaluations. In contrast, reverse KL minimization requires re-evaluating the model at new quadrature points every time the transport map changes (since the reverse KL integrates over the changing pullback distribution instead of over the static target).
Also, the cross-entropy loss function $\mathcal{L}$ is well-known to be \textit{moment-matching}, so any minimizer $\surrogate(\cdot;\bfc^{*})$ will tend to have similar expectations as $\poster$. On the other hand, reverse KL minimization is \textit{mode-seeking},
which generally yields poor approximations of multi-modal targets~\cite{felardosDesigningLossesDatafree2023} and has been described as inadequate for importance sampling~\cite{jerfelVariationalRefinementImportance2021}.

An alternative to variational approximation of transports is given by the tensor-based transport methods of \cite{dolgovApproximationSamplingMultivariate2020,cuiDeepCompositionTensorTrains2022,cuiScalableConditionalDeep2023,cuiDeepImportanceSampling2024}, %
which \textit{interpolate} the unnormalized target density and then explicitly approximate the associated Knothe--Rosenblatt transport map via tensor-based integration. Some of these methods also employ a sequence of bridging densities. See~\cite{cuiDeepImportanceSampling2024} for a comparison to cross-entropy methods. \textit{Deep} tensor methods are necessarily constructed as a composition of maps. Empirically, we observe numerical instability when composing more than a few maps, i.e., choosing $\eta_{\stepIDX+1} = \surrogate_\stepIDX$ for more than one or two consecutive steps. This difference may be due to the interpolation property of tensor methods, compared to our variational procedure.

To our knowledge, no theory characterizes the convergence of $Q_\stepIDX^\MIS$ in \cref{alg:multiple_IS}, particularly when these quadrature rules are employed iteratively in the context of a cross-entropy method.
Nevertheless, the re-use of expensive likelihood evaluations is crucial in the small-sample setting of interest here.
Our weighting of different biasing distributions as in \cref{alg:multiple_IS} extends the unweighted mixing of datasets suggested in contemporaneous work~\cite{wangMitigatingModeCollapse2025,kimSequentialNeuralJoint2025}. In particular, the partition of unity $\alpha_j$ in our scheme makes the different transport maps mix according to how well they ``trust'' one another, i.e., how close the surrogate distributions are to one another when evaluated at the quadrature points. %
This mixing emulates characteristics of defensive importance sampling~\cite{owen2000safe,hesterberg1995weighted} as well as some behavior of \textit{contrastive learning}~\cite{baptistaMathematicalPerspectiveContrastive2025}. Samples in  $Q_\stepIDX^\MIS$ not only help the transport map approximate the posterior where it is concentrated but also reveal where little posterior mass is present.

\section{Numerical examples}  \label{sec:examples}

All forward models are implemented in the open-source \texttt{MrHyDE} toolkit~\cite{wildeyMrHyDEV102023} and transport maps using the \texttt{MParT} package~\cite{parnoMParTMonotoneParameterization2022}.%
\footnote{Code to reproduce all examples is available at \url{https://github.com/dannys4/MultilevelTransport}.} %
Our forward models are induced by various parametric partial differential equations (PDEs), each described in the ensuing subsections. The parameters $\theta$ of each PDE are endowed with a uniform prior on the \ymmtext{$d$-dimensional} hypercube
${[0,1]}^{d}$, for values of $d$ specified below. To integrate over this uniform distribution, we use the randomized QMC method described in \Cref{app:qmc}. The likelihood function is of the form \eqref{eqn:gaussian_obs} with isotropic Gaussian observation errors. In particular, we define a hierarchy of forward models ${(G_\fidelity)}_{\fidelity=0}^{L}$ with $G_\fidelity:[0,1]^d\to\RR^m$ %
and
$\likelyFidel(\theta) \propto \exp(-\frac{1}{2\sigma^{2}}  \|G_{\fidelity}(\theta) - \bfy^{*}  \|^{2})$
with $\prior(\theta) \equiv 1$.
Choosing target model $G_{L}$ gives $\likely_{L}$ as our highest fidelity likelihood, so $\fidelity_{M}= L$ and $\fidelity_{0}= 0$.

The PDEs in the examples below are all defined on the domain $\PDEDomain=[0,1]^2$ (where this spatial dimension should not be confused with
the parameter dimension $d$). The data $\bfy^{*}  \in\RR^{m}$ correspond to observations of the PDE solution at $m$ points in $\PDEDomain$. To avoid an ``inverse crime''~\cite{kaipioStatisticalInverseProblems2007}, we generate $\bfy^{*}$ from a higher-fidelity PDE solution than $G_{L}$ (i.e., a finer spatial/temporal discretization) and, unless otherwise stated, add noise to these observations.
In our multi-fidelity examples below, we use tempering thresholds $(t_{1},t_{2}) = (0.5, 0.8)$,
i.e., increase the fidelity when $\beta_{j-1}  \geq 0.5$ and again when $\beta_{j-1}  \geq
0.8$. We denote the number of new model evaluations at step $\stepIDX$ by
$n_{\stepIDX}$ and the parameter count in the biasing distribution $\surrogateStep$, i.e., the dimension of $\FF_\stepIDX$, by $p_{\stepIDX}$. \ymmtext{Overall, we seek to use these four examples to demonstrate the different characteristics we seek for Bayesian computational approaches: accurate posterior density approximation without gradients, %
robustness to unusual concentration and mode separation, resilience to poor intermediate approximations, and a balance between exploration and exploitation for quadrature with a limited budget.}

\subsection{Error metrics}
We quantify the error at step $\stepIDX$ four ways. To assess
the error in each quadrature rule $Q_\stepIDX^\MIS$ relative to $\poster_M$,
we report the
error in the mean (RMSE), defined as $D(\poster_M, \surrogate_\stepIDX) = \|\EE_{\poster_M}[\Theta] - \widehat{\mu}(Q_\stepIDX^\MIS)\|_2$, where $\widehat{\mu}(Q) = \sum_{k=1}^N w^{(k)}  \theta^{(k)}$ for a quadrature rule $Q=(\theta^{(k)},w^{(k)})_{k=1}^N$.
Similarly, we compute the covariance matrix of $\Theta$ using $Q_\stepIDX^\MIS$ and report the F\"{o}rstner distance~\cite{forstnerMetricCovarianceMatrices2003} (the geodesic distance on the symmetric positive-definite manifold) between this approximation and the true covariance.
To quantify %
the discrepancy between surrogate density $\surrogateStep$ and the target high-fidelity posterior $\poster_{M}$, we %
estimate the maximum mean discrepancy ($\MMD$)~\cite{muandetKernelMeanEmbedding2017}
between the two, with the squared-exponential and Mat\`{e}rn-1.5 kernels; these errors are denoted $\MMD_{\textrm{G}}$ and $\MMD_{\textrm{M}_{1.5}}$, respectively. Thus, we quantify error of $Q_\stepIDX^\MIS$ in two ways and measure the error of $\surrogateStep$ in two ways.
For each of these four error discrepancies, labeled $D(\cdot, \cdot)$, we report $D(\surrogateStep,\poster_{M}) / D(\prior, \poster_{M})$. This relative scaling to ensures that, even if the prior is a good approximation to the posterior,
we still improve on it. See \ymmtext{the supplementary material's} \Cref{sm:error_metrics} for a more thorough discussion of our error metrics, including the computation of ``reference'' values of the posterior moments.

\subsection{Diffusion single-source inversion}
\label{sec:diffusion_single} 
\begin{figure}[!t]
	\centering
	\includegraphics[width=0.7\textwidth, trim={0 1.1cm 0 1.5cm}]{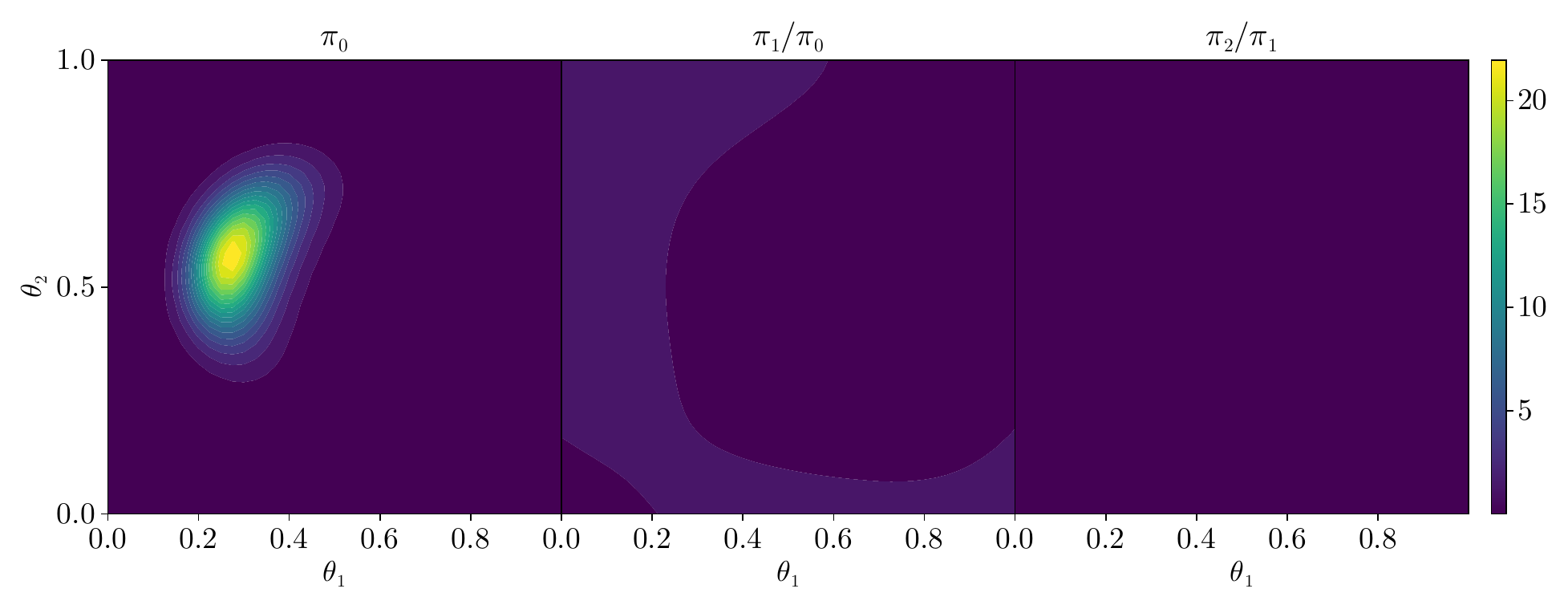}%
	\caption{Likelihood ratios induced by the three PDE discretizations in \Cref{sec:diffusion_single}.}%
	\label{fig:likelihood_ratios}
\end{figure}

As in~\cite{farcasMultilevelAdaptiveSparse2020}, we start with the elliptic PDE
\begin{equation}
	\label{eqn:diffusion_single_pde}  \Delta u = A(\alpha)\exp(-\frac{1}{2\alpha^{2}}  \|\bfx - \theta\|^{2}
	),\quad u(\mathbf{x}) \equiv 0,\ \mathbf{x}  \in\partial\PDEDomain,
\end{equation}
where $A(\alpha) = \frac{5}{\tau\alpha}$ and $\alpha=0.15$, with $\tau$ twice
Archimedes' constant. This is a simple example with a linear PDE and $d=2$,
but nonetheless offers algorithmic insight due to its concentrated non-Gaussian posterior.
We use 16 observations
at $(0.2i,0.2j)$ for $i,j \in \{1,2,3,4\}$, with noise variance $\sigma^2 = 0.04$. We define our models using successively finer meshes in a finite-element solver. For more details on discretizations and solvers in all examples, see \Cref{sm:fem_discretization}.
In \Cref{fig:likelihood_ratios}, we compare the likelihoods $\likely_0$, $\likely_1$, $\likely_2$ and see diminishing returns as we increase fidelity; $\likely_1$ is quite close to $\likely_2$.

\begin{figure}[ht!]
	\centering
	\includegraphics[clip, trim={0 0.25cm 0 0}, width=0.8\linewidth]{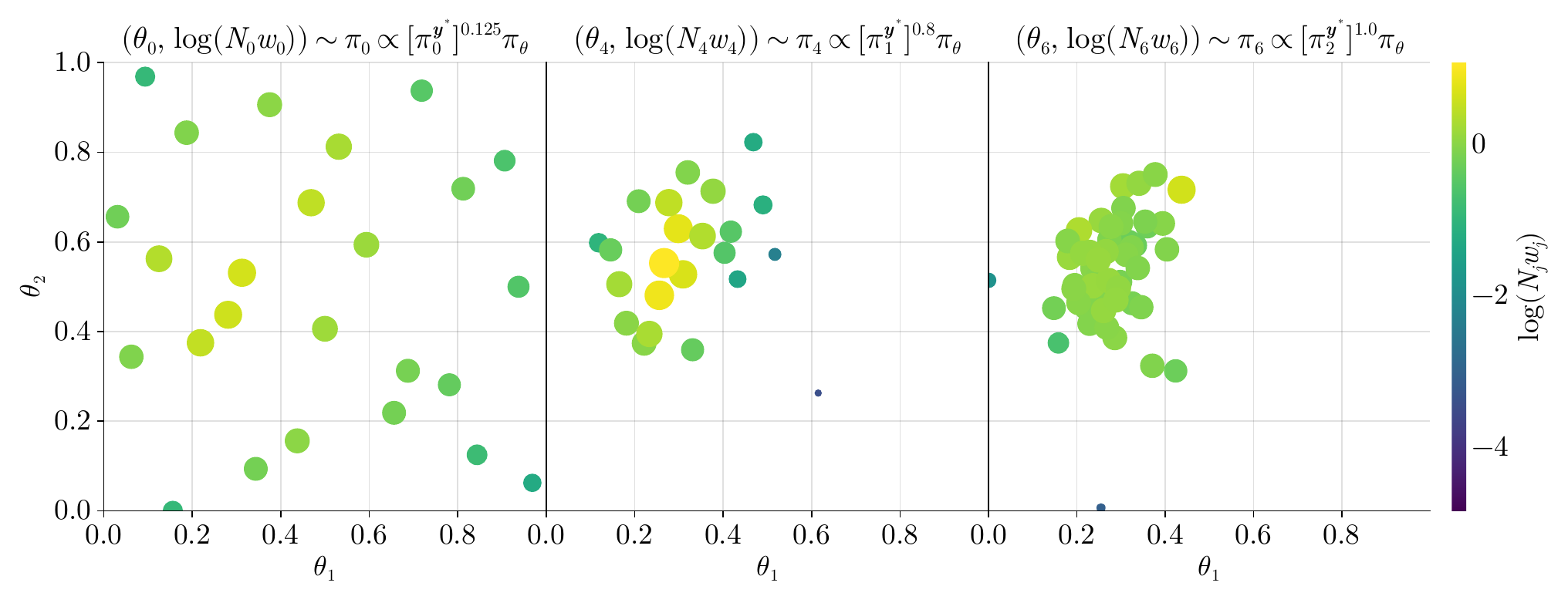}
	\caption{Quadrature rules $Q_\stepIDX^\MIS$ for \Cref{sec:diffusion_single}. Both color and size reflect the weight of a point, with uniform weights desired.}%
	\label{fig:diffusion_single_quad}
\end{figure}

\begin{figure}[!ht]
	\centering
	\includegraphics[clip, trim={0 0.3cm 0 0.6cm}, width=0.65\linewidth]{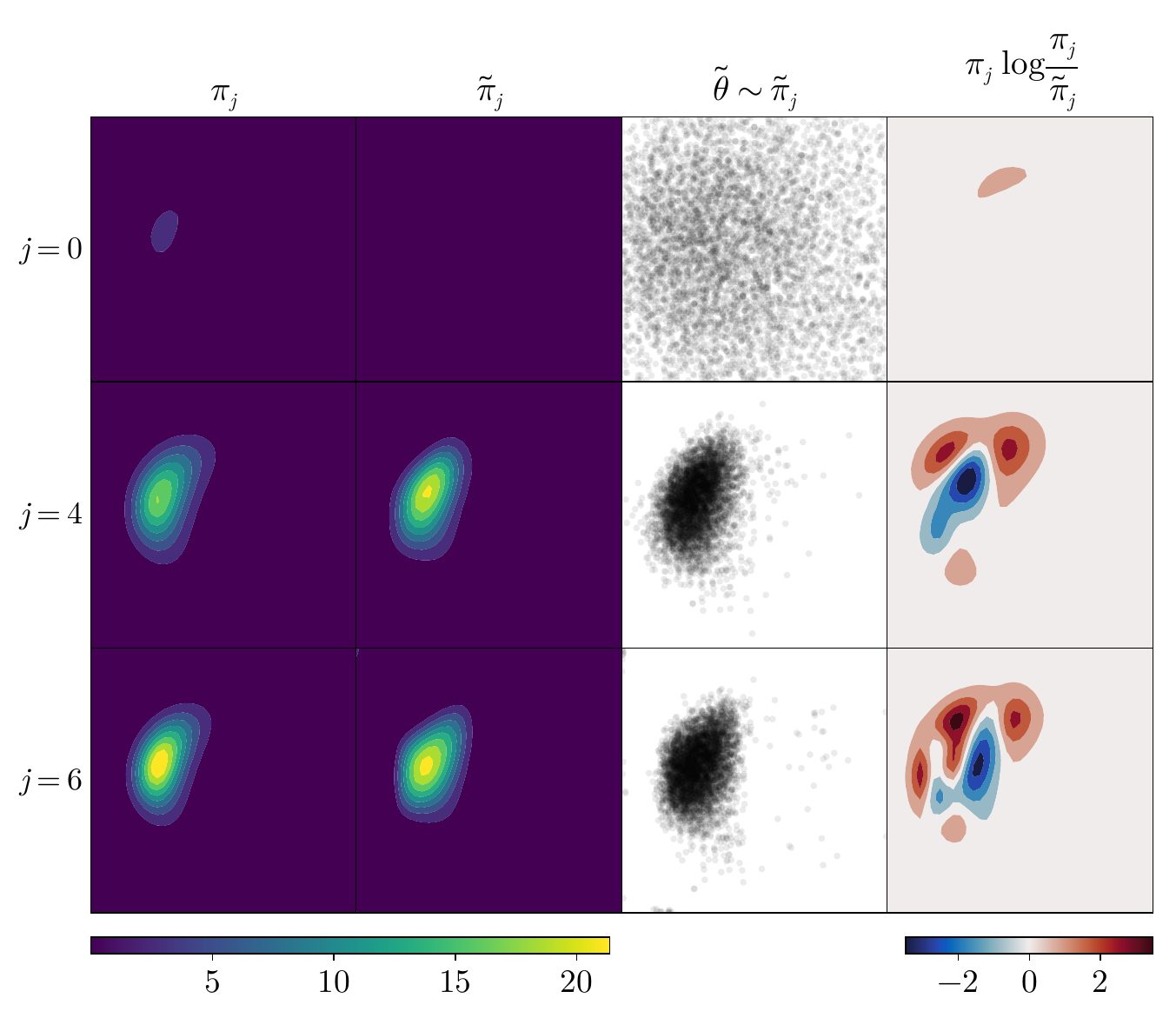}
	\caption{Visualizing the results for \Cref{sec:diffusion_single}. Columns from left to right: target density $\poster_j$, transport-induced surrogate density $\surrogate_\stepIDX$, samples from the surrogate, and weighted log-ratio of $\poster_j$ to $\surrogate_\stepIDX$.}%
	\label{fig:diffusion_single_pullbacks}
\end{figure}

\Cref{tab:diffusion_single_results} shows hyperparameters and numerical results of running our algorithm. Visualizations of $\posterStep$, $\surrogateStep$, and $\posterStep\log\frac{\posterStep}{\surrogateStep}$
(i.e., the integrand in the KL divergence) are given in \Cref{fig:diffusion_single_pullbacks}. \ymmtext{We see that the surrogate density and samples closely match the desired target}. We also show, in \Cref{fig:diffusion_single_quad}, a few of the quadrature rules $Q_j^\MIS$ produced by the algorithm. \ymmtext{Recall the discussion of reference choice in \cref{sec:reference}; the interplay between the reference $\eta_j$ and family of densities $\FF$ can be seen in the choice of $p_j$, where we choose a smaller $p_j$ on the steps that $\ell$ increases. We re-emphasize that not only do we capture quantities of scientific interest (reflected by the RMSE in the mean and F\"{o}rstner error in the covariance), but our overall error decreases continually with a small evaluation budget.} This example demonstrates that we can efficiently and accurately capture the target density
using \Cref{alg:core_loop} with only $175$ model evaluations: simulating the low,
medium, and high-fidelity models $100$, $25$, and $50$ times, respectively.
\begin{table}[ht!]
	\centering
    \caption{Hyperparameters and error metrics for \Cref{sec:diffusion_single}. Row $\stepIDX$ corresponds to $Q_{\stepIDX}^\MIS$ and $\surrogateStep$. $n_\stepIDX = 25$ at each step.}%
	\begin{tabular}{@{}c cccc cccc@{}}
		\toprule	& \multicolumn{3}{c}{Hyperparameters} & & \multicolumn{4}{c}{Error Metrics}  \\
		\cmidrule(rl){2-4}  \cmidrule(rl){6-9} $\stepIDX$ & $\fidelityStep$ & $\temperStep$ & $p_{\stepIDX}$ & $\rESS$ & RMSE & F\"{o}rstner & $\MMD_{\mathrm{M}_{1.5}}$ & $\MMD_{\mathrm{G}}$ \\ \midrule
		0 & 1 & 0.125 & 14 & 8.20\E{-1} & 6.78\E{-1} & 8.94\E{-1} & 8.88\E{-1} & 8.93\E{-1} \\
		1 & 1 & 0.250 & 14 & 6.61\E{-1} & 4.70\E{-1} & 7.15\E{-1} & 7.76\E{-1} & 7.86\E{-1} \\
		2 & 1 & 0.400 & 20 & 5.36\E{-1} & 2.81\E{-1} & 5.29\E{-1} & 6.32\E{-1} & 6.44\E{-1} \\
		3 & 1 & 0.500 & 20 & 5.19\E{-1} & 2.21\E{-1} & 4.19\E{-1} & 5.41\E{-1} & 5.54\E{-1} \\
		4 & 2 & 0.800 & 9  & 6.06\E{-1} & 3.12\E{-2} & 4.62\E{-2} & 8.17\E{-2} & 8.12\E{-2} \\
		5 & 3 & 1.000 & 9  & 9.50\E{-1} & 8.00\E{-2} & 6.00\E{-2} & 1.65\E{-1} & 1.69\E{-1} \\
		6 & 3 & 1.000 & 35 & 9.31\E{-1} & 6.33\E{-2} & 7.88\E{-2} & 1.38\E{-1} & 1.40\E{-1} \\
		\bottomrule
	\end{tabular}%
	\label{tab:diffusion_single_results}
\end{table}

\subsection{Diffusion multi-source inversion}
\label{sec:diffusion_multi} 
\begin{figure}[!ht]
	\centering
	\includegraphics[clip, trim={0 0.5cm 0.25cm 0.45cm}, width=0.6\textwidth]{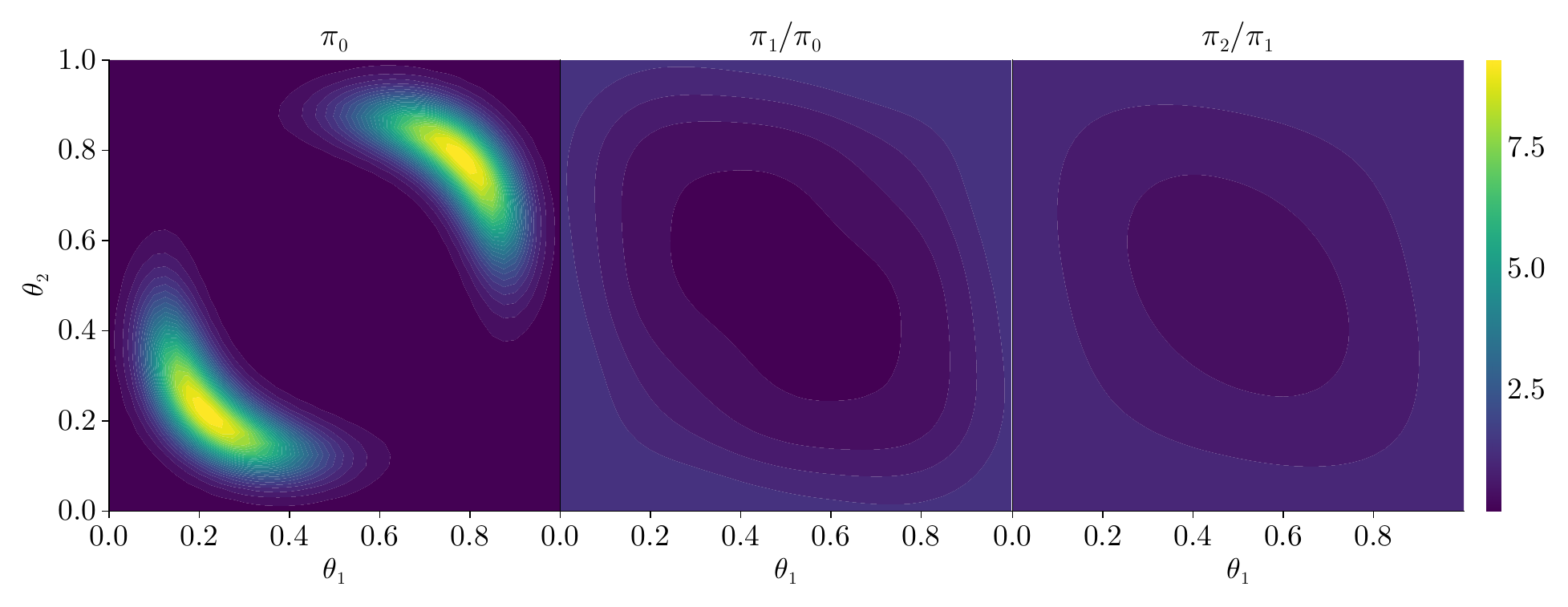}
	\caption{Likelihood ratios induced by the three PDE discretizations in \Cref{sec:diffusion_multi}.}%
	\label{fig:likelihood_ratios_multisource}
\end{figure}

\begin{figure}[ht!]
	\centering
	\includegraphics[clip, trim={0 0.5cm 0 0.5cm}, width=0.8\linewidth]{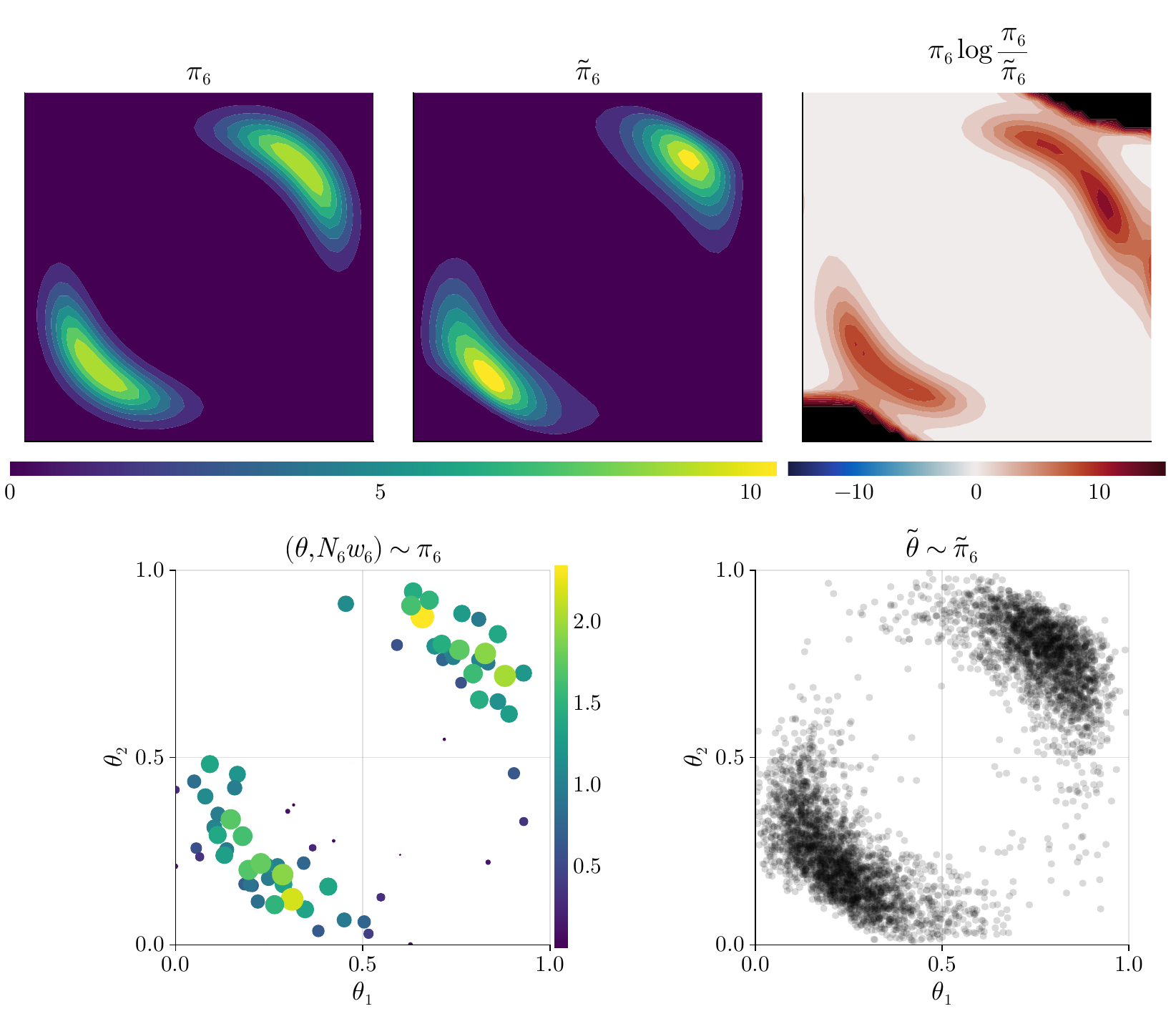}
	\caption{Visualizing the results for \Cref{sec:diffusion_multi}. (Top): From left
	to right, we compare the truth, surrogate, and KL integrand (where a black region for the KL integrand corresponds to when $\surrogate_M(\theta) = 0$ numerically). (Bottom left): Final quadrature $Q_M^\MIS$. (Bottom right): Samples from the pullback distribution $\surrogate_{M}$.}%
	\label{fig:diffusion_multi_results}
\end{figure}

We now tackle the bimodal experiment from~\cite{farcasMultilevelAdaptiveSparse2020}
\begin{equation}
	\begin{gathered}
		\Delta u = A(\alpha)\exp(-\frac{1}{2\alpha^{2}}  \|\bfx-\theta\|^2)+ b A(\alpha)\exp(-\frac{1}{2\alpha^{2}}  \|\bfx-\nu\|^2);\ \ u(\mathbf{x})
		\equiv 0,\;\mathbf{x}  \in\partial\PDEDomain,
	\end{gathered}
\end{equation}
where $b \in \{0,1\}$ dictates whether we evaluate the PDE or generate data, respectively: we use two localized sources to generate data but a single-source parameterization of the right-hand side (cf. \Cref{sec:diffusion_single}) when performing inference. The data is generated
using $\theta=(0.15,0.15)$ and $\nu=(0.85,0.85)$. To preserve strong bimodality and interesting
features of the target density, we add no observation noise %
to generate $\bfy$,
observed at the same sensor locations as \Cref{sec:diffusion_single}.
Ratios of the successive likelihoods are visualized in \Cref{fig:likelihood_ratios_multisource}.
As the prior and true posterior have identical mean $(0.5,0.5)$, we report the average of $Q^{\MIS}_\stepIDX$ instead of RMSE. Quantitative results of our algorithm are in \Cref{tab:diffusion_multi_results} and the final approximation $\surrogate_{M}$ of the target distribution is visualized in \Cref{fig:diffusion_multi_results}. The column of \Cref{tab:diffusion_multi_results} labeled $\tilde{\mu}_j$ shows the quadrature approximation of the posterior mean via $Q_j^\MIS$ at each step; these values should be compared to the true mean, $(0.5, 0.5)$.

\begin{table}[ht!]
	\centering
	\caption{Hyperparameters and error metrics for \Cref{sec:diffusion_multi}. Row $\stepIDX$ corresponds to $Q_{\stepIDX}^\MIS$ and $\surrogateStep$.}%
	\begin{tabular}{@{}c ccc c cc ccc@{}}
		\toprule	& \multicolumn{3}{c}{Hyperparameters} & & & & \multicolumn{3}{c}{Error Metrics}  \\
		\cmidrule(rl){2-4}\cmidrule(rl){8-10}
		$j$ & $\ell_j$ & $\beta_j$ & $p_j$ & $\mathrm{rESS}$ & \multicolumn{2}{c}{$\tilde{\mu}_\stepIDX$} & F\"{o}rstner & $\MMD_{\mathrm{M}_{1.5}}$ & $\MMD_{\mathrm{G}}$ \\
		\midrule
		0 & 1 & 0.200 & 14 & 8.21\E{-1} & 0.487 & 0.504 & 8.99\E{-1} & 9.63\E{-1} & 9.66\E{-1} \\
		1 & 1 & 0.325 & 27 & 6.63\E{-1} & 0.476 & 0.488 & 6.58\E{-1} & 6.38\E{-1} & 6.46\E{-1} \\
		2 & 1 & 0.500 & 27 & 5.38\E{-1} & 0.462 & 0.499 & 4.86\E{-1} & 4.95\E{-1} & 5.02\E{-1} \\
		3 & 2 & 0.800 & 9  & 6.27\E{-1} & 0.521 & 0.526 & 4.55\E{-2} & 4.51\E{-1} & 4.59\E{-1} \\
		4 & 3 & 1.000 & 9  & 7.18\E{-1} & 0.466 & 0.479 & 4.16\E{-2} & 4.23\E{-1} & 4.32\E{-1} \\
		5 & 3 & 1.000 & 27 & 6.83\E{-1} & 0.469 & 0.469 & 4.16\E{-2} & 2.41\E{-1} & 2.45\E{-1} \\
		6 & 3 & 1.000 & 35 & 7.52\E{-1} & 0.470 & 0.481 & 4.30\E{-2} & 1.94\E{-1} & 1.95\E{-1} \\
	\bottomrule
	\end{tabular}
	\label{tab:diffusion_multi_results}
\end{table}

The performance of our algorithm improves significantly on results from~\cite{farcasMultilevelAdaptiveSparse2020}; we capture multimodal behavior with only $p_M=35$ parameters in the density surrogate $\surrogate_M$.
By contrast, \cite{farcasMultilevelAdaptiveSparse2020}~fails to accurately capture multi-modality even with $O(100)$ parameters in its interpolant, demonstrating that the dataset used to construct the surrogate (here, our $Q^{\MIS}_M$, in the bottom left of \Cref{fig:diffusion_multi_results}) is the crucial factor, not the complexity of the surrogate itself. %
In this example, we see a continual decrease in errors; in fact, the final MMD and covariance  errors relative to the high-fidelity posterior are on the order of $10^{-1}$, but we only use 75 high-fidelity posterior evaluations and 175 total model evaluations. Errors in the mean (compared to the true value) are on the order of $10^{-2}$.
For comparison, if we could hypothetically draw 75 independent and unweighted Monte Carlo samples from $\poster_{M}$ directly (which of course is not possible), {the resulting Monte Carlo standard error in the mean estimate would be approximately 3.42\E{-2}. Thus, our algorithm performs comparably to or better than perfect Monte Carlo sampling of $\poster_M$; compare this to the asymptotic bias in \cite[Figure 7]{farcasMultilevelAdaptiveSparse2020}.}%

\subsection{CDR initial condition inversion}
\label{sec:cdr} 
\begin{figure}[ht!]
	\centering
	\includegraphics[clip, trim={0 0.5cm 0 0.5cm}, width=0.8\linewidth]{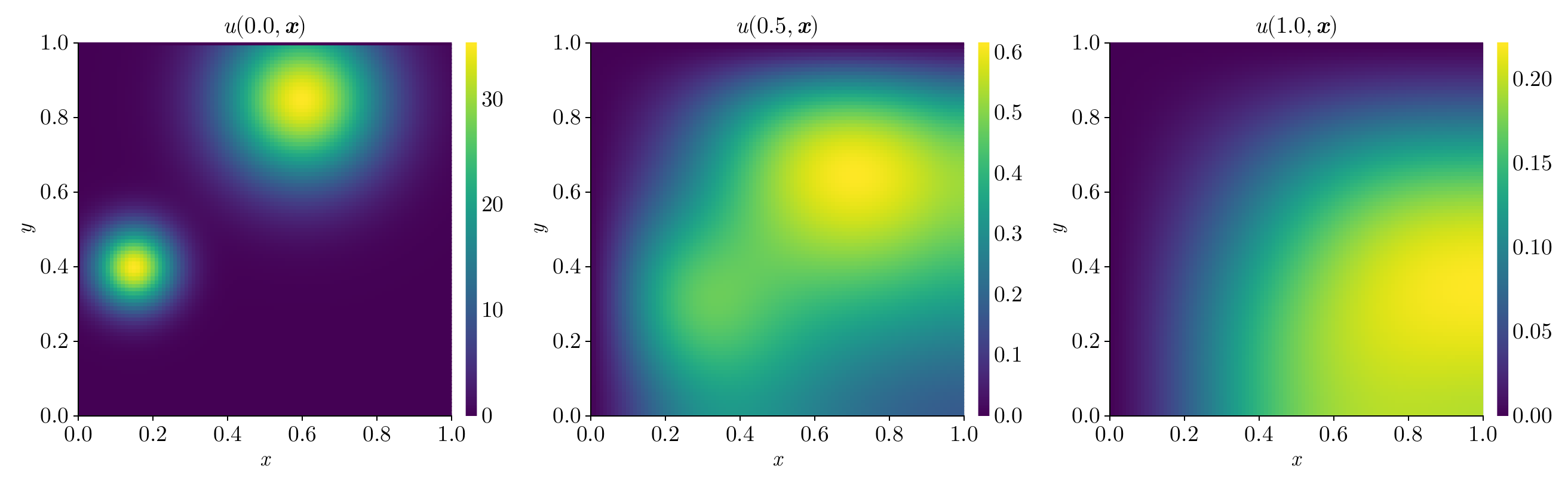}
	\caption{Solution to PDE~\eqref{eqn:cdr} for given convection $\mathbf{c}^{\mathrm{true}} = \mathbf{c}_{2}=(0.5,-0.5)$ at time
	snapshots $t=0,0.5,1$.}%
	\label{fig:cdr_state}
\end{figure}

\begin{figure}[ht!]
	\centering
	\includegraphics[width=0.9\linewidth]{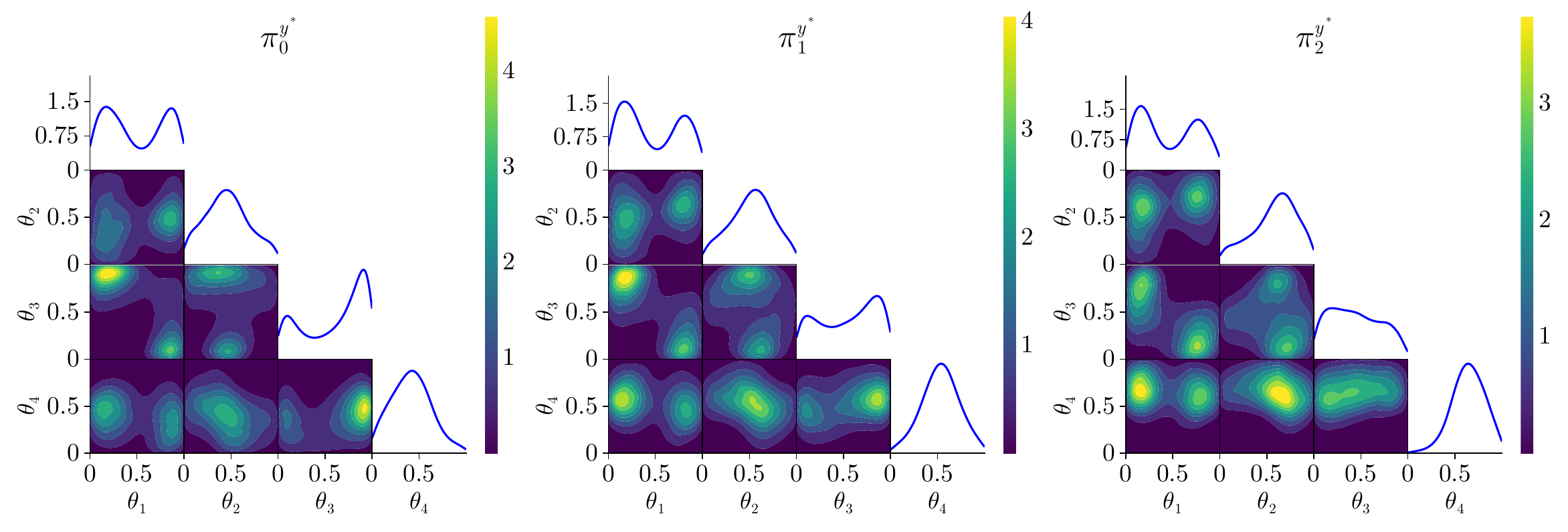}
	\caption{Kernel density estimators (KDEs) of likelihood marginals in \Cref{sec:cdr} for each fidelity.
	}%
	\label{fig:cdr_densities}
\end{figure}

Consider the convection-diffusion-reaction PDE,
\begin{equation}
	\label{eqn:cdr}
	\begin{gathered}
		\frac{\partial u}{\partial t} - \nabla\cdot (\kappa\nabla u) -\mathbf{c}_\fidelity \cdot \nabla
		u - 5u^2 = 0\;\;\; \forall\mathbf{x}  \in \PDEDomain,\ t\in(0,10];\quad u(t,\mathbf{x}) \equiv 0\;\;\;\forall \mathbf{x} \in\partial\PDEDomain,\ t\in(0,10]\\
        u(0,\mathbf{x}) = A(\alpha)\left[\exp\left(-\frac{4}{2\alpha^{2}} \|\mathbf{x}-\mathbf{x}_1\|^2\right)+\exp\left(-\frac{1}{2\alpha^{2}} \|\mathbf{x}-\mathbf{x}_2\|^2\right)\right]\;\;\;\forall\mathbf{x} \in \PDEDomain.
	\end{gathered}
\end{equation}
We use zero-Dirichlet conditions on the top and left boundaries and zero-Neumann conditions on the right and bottom boundaries. We wish to infer the initial condition of this nonlinear and time-dependent PDE, parameterized by $\theta=(\bfx_{1},\bfx_{2}) \in [0,1]^4$,
from observations that are sparse in both space and time. The data-generating parameters are $\mathbf{x}^{\mathrm{true}}_{1}=(0.6,0.85)$
and $\mathbf{x}^{\mathrm{true}}_{2}=(0.15,0.40)$. Here, we keep constant mesh sizes across the
levels $\stepIDX$ and instead model the Peclet number as increasing across fidelities $\fidelity$. We impose this using velocity vectors of increasing magnitude, $\mathbf{c}_{\fidelity} \equiv \frac{\fidelity}{2}(1,-1)$ for $\fidelity=0,1,2$, where data for inference are generated at $\mathbf{c}^{\mathrm{true}} \equiv \mathbf{c}_{2}$. Our observations are taken at the same spatial locations as in the previous two examples, tensorized with six observation times $t \in \{0.5,0.6,\ldots,1\}$. The PDE solution $u$ obtained with the true convective term $\mathbf{c}_{2}$ is pictured in \Cref{fig:cdr_state}. In \Cref{fig:cdr_densities}, we visualize one- and two-dimensional marginals of the target distributions $\poster_0$, $\poster_1$, and $\poster_2$. \ymmtext{These plots demonstrate that the modes deform nonlinearly as the convection speed increases.}

\begin{table}[ht!]
	\centering
	\caption{Hyperparameters and error metrics for \Cref{sec:cdr}. Row $\stepIDX$ corresponds to $Q_{\stepIDX}^\MIS$ and $\surrogateStep$. $n_\stepIDX = 75$ at each step.}%
	\begin{tabular}{@{}c ccc ccccc@{}}
		\toprule	& \multicolumn{3}{c}{Hyperparameters} & & \multicolumn{4}{c}{Error Metrics}  \\
		\cmidrule(rl){2-4}  \cmidrule(rl){6-9}
		$\stepIDX$	& $\fidelityStep$ & $\temperStep$ & $p_{\stepIDX}$ & $\rESS$ & RMSE & F\"{o}rstner & $\MMD_{\mathrm{M}_{1.5}}$ & $\MMD_{\mathrm{G}}$ \\
		\midrule
			0 & 1 & 0.300 & 34  & 8.22\E{-1} & 1.24\E{ 0} & 8.19\E{-1} & 1.01\E{ 0} & 1.01\E{ 0} \\
			1 & 1 & 0.500 & 125 & 6.64\E{-1} & 1.49\E{ 0} & 6.67\E{-1} & 1.11\E{ 0} & 1.11\E{ 0} \\
			2 & 2 & 0.675 & 69  & 5.33\E{-1} & 9.06\E{-1} & 6.12\E{-1} & 9.47\E{-1} & 9.69\E{-1} \\
			3 & 2 & 0.800 & 125 & 5.02\E{-1} & 9.67\E{-1} & 4.38\E{-1} & 9.09\E{-1} & 9.34\E{-1} \\
			4 & 3 & 0.800 & 34  & 2.29\E{-1} & 3.93\E{-1} & 8.04\E{-1} & 7.44\E{-1} & 7.87\E{-1} \\
			5 & 3 & 0.800 & 125 & 4.05\E{-2} & 1.37\E{ 0} & 1.09\E{ 0} & 2.09\E{ 0} & 2.33\E{ 0} \\
			6 & 3 & 0.800 & 329 & 3.15\E{-1} & 5.21\E{-1} & 8.34\E{-1} & 7.72\E{-1} & 8.38\E{-1} \\
			7 & 3 & 0.800 & 329 & 2.64\E{-1} & 4.24\E{-1} & 7.10\E{-1} & 7.86\E{-1} & 8.49\E{-1} \\
			8 & 3 & 1.000 & 329 & 5.66\E{-1} & 3.79\E{-1} & 7.87\E{-1} & 7.89\E{-1} & 8.46\E{-1} \\
		\bottomrule
	\end{tabular}
	\label{tab:cdr_params}
\end{table}

\begin{figure}
	\centering
	\includegraphics[width=0.35\linewidth]{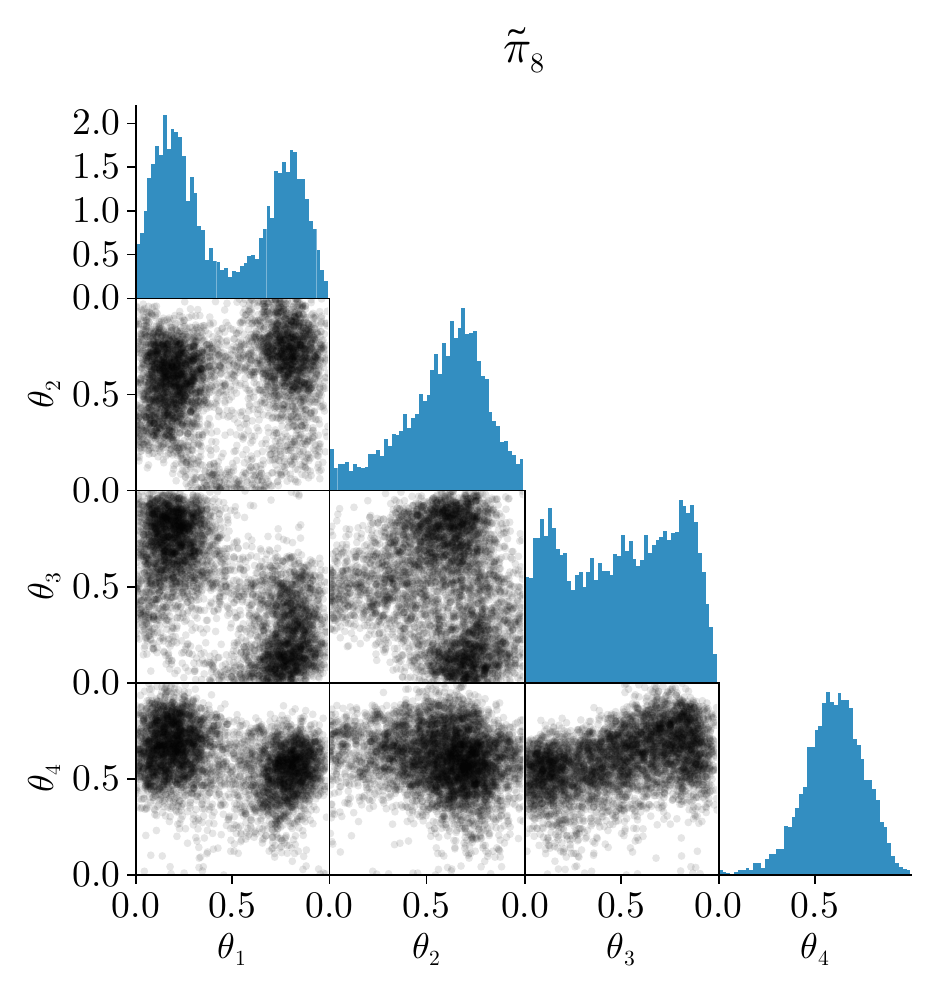}
	\includegraphics[width=0.35\linewidth]{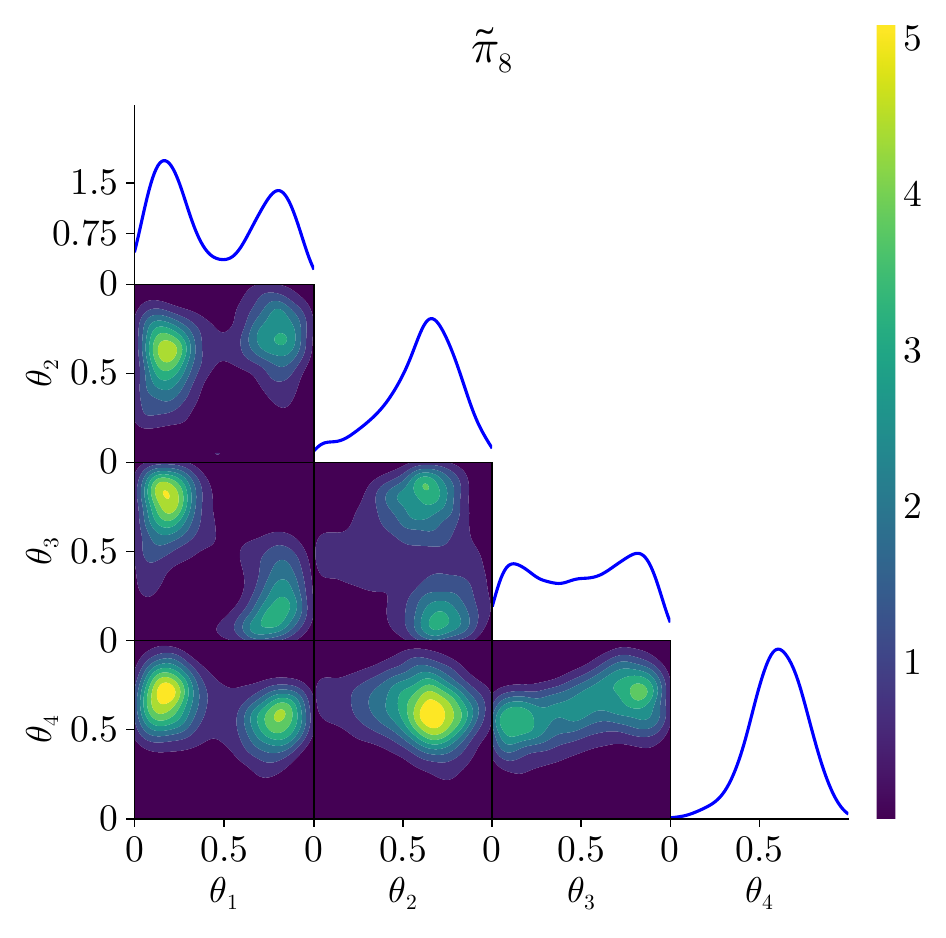}
	\caption{Visualization of the results for \Cref{sec:cdr}.
	(Left): samples $\widetilde\theta\sim\surrogate_{8}$ with one-dimensional histograms on the diagonal. (Right): KDE of
	the two-dimensional marginals of $\surrogate_{8}$ with one-dimensional KDE on the diagonal.}%
	\label{fig:cdr_pullback}
\end{figure}
Hyperparameter values for our algorithm and error metrics over successive iterations are reported \Cref{tab:cdr_params}, where we choose $n_\stepIDX = 75$ for this higher-dimensional example. Samples and marginal densities of $\surrogate_{M}$ are shown in \Cref{fig:cdr_pullback}.
Because the errors in \Cref{tab:cdr_params} are relative, their scale is not affected by the parameter dimension $d$.
We do see slightly slower reduction of the error with the number of samples compared to the prior examples, which is expected for several reasons. The larger parameter dimension $d$ (i) leads to larger errors in our quadrature approximation for the same number of model evaluations,
and (ii) requires searching within larger classes of functions $(\FF_\stepIDX)_\stepIDX$ for our transport maps. Additionally---by construction---this example contains significantly more model mismatch
between successive fidelities, compared to the previous examples. Thus when we
transition from an intermediate Peclet number model to the final model at step $j=4$, our biasing distribution proposes inefficient and somewhat uninformative samples. Nevertheless, we are able to correct for the bias in the low-fidelity models and efficiently capture the final target distribution's multi-modality and anisotropy with relatively few evaluations: compare the rightmost axes in \Cref{fig:cdr_densities,fig:cdr_pullback}.

The reduction of error from step $\stepIDX=5$ to $\stepIDX=6$ demonstrates robustness to poor intermediate surrogates, due to our multiple IS procedure: the weighting step in multiple IS allows us to recognize the poor efficiency of the rule $Q_5$ and defensively rely on samples from steps $\stepIDX=4$ and $\stepIDX=6$. We find that this reweighting approach, not present in \cite{wangMitigatingModeCollapse2025,kimSequentialNeuralJoint2025}, prevents the algorithm from diverging due to poor intermediate approximations, and is vital to Bayesian inference with small model evaluation budgets and in multi-fidelity settings where differences between available models may not be well-controlled.

\subsection{Helmholtz inversion}
\label{sec:helmholtz} 
\begin{figure}[ht!]
	\centering
	\includegraphics[clip, trim={0 0.85cm 0 0.19cm}, width=0.35\linewidth]{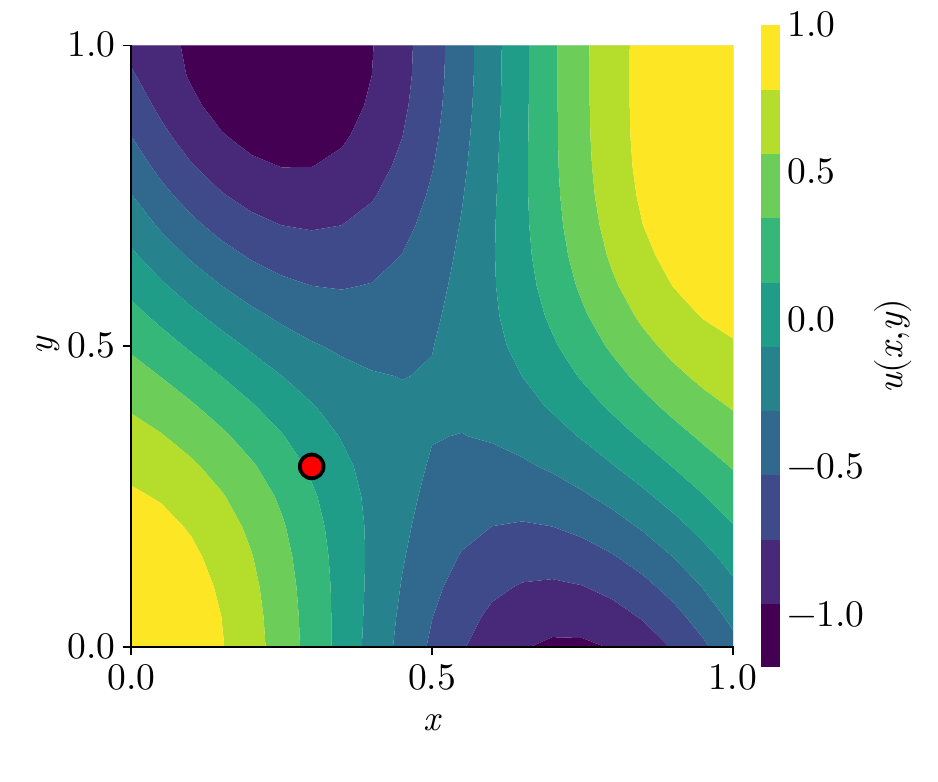}
	\caption{True solution of PDE \eqref{eqn:helmholtz_pde}, where the red point is the location $\theta^{\mathrm{true}}$.}%
	\label{fig:helmholtz_state}
\end{figure}

\begin{figure}[ht!]
	\centering
	\includegraphics[width=0.7\linewidth]{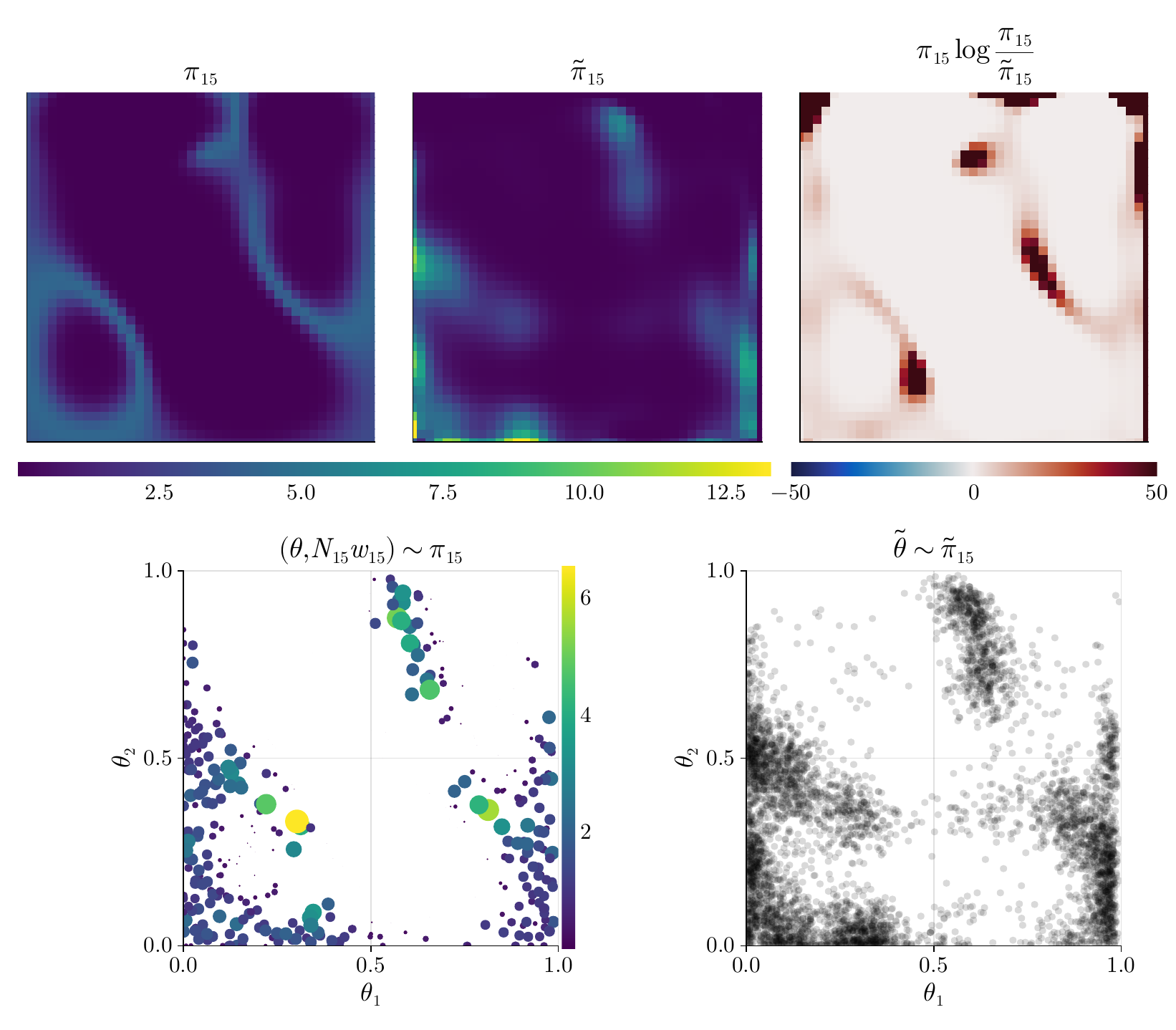}
	\caption{Visualization of the results for \Cref{sec:helmholtz}.
	(Top): From left to right, we compare the truth, surrogate, and KL integrand. (Bottom left): Final quadrature $Q_M^\MIS$. (Bottom right): Samples from the pullback distribution $\surrogate_{M}$.
	}%
	\label{fig:helmholtz_results}
\end{figure}

This problem is motivated by anomalous behavior in a dielectric medium excited by a signal field~\cite{tadiInverseProblemHelmholtz2011}. Suppose we have a wave simulated by solving a Helmholtz equation with a point-disturbance located at $\theta$; our PDE is parameterized as
\begin{gather}\label{eqn:helmholtz_pde}
    k^{2}(1+\exp(-\frac{1}{2\alpha^{2}}  \|\mathbf{x}-\theta\|^{4}_2))u(\mathbf{x}) + \nabla^{2}u(\mathbf{x}) = 0,\quad \mathbf{x}  \in \PDEDomain \\
    u(x,0) = \cos(k\ x\ \cos(\gamma)),\ u(0,y) = \cos(k\ y\ \sin(\gamma)),\\
    u(1,y) = \cos(k(\cos(\gamma)+y\sin(\gamma))),\ \frac{\partial u}{\partial y}(x,1) \equiv 0
\end{gather}
Here, we observe the PDE solution (with noise) at the two locations $(0.33,1)$ and $(0.66,1)$, e.g., we can only observe the behavior of the signal %
on the boundary of the domain. Further, we fix wave speed $k=5$, wave angle $\gamma=\frac{\tau}{12}$, and anomaly concentration $2\alpha^{2}=10^{-4}$.
The data were generated with parameter value $\theta^{\mathrm{true}}=(0.3,0.3)$.
The solution to PDE \eqref{eqn:helmholtz_pde} associated with $\theta^{\mathrm{true}}$ %
is pictured in \Cref{fig:helmholtz_state}.
To show the efficacy of \Cref{alg:multiple_IS} in conjunction with tempering, %
we consider only a single fidelity (i.e., $L=0$)
and use a small noise variance $\sigma^{2}= 0.1\times 2^{-6}$.
The elliptic form of the PDE, small observational noise, and sparse observations make the posterior $\poster$ have well-separated, nonlinear ridges with significant concentration. Results are reported in \Cref{tab:helmholtz_params} and \Cref{fig:helmholtz_results}.

This exercise demonstrates that tempering enables the surrogate (and its associated quadrature rule) to focus on regions of posterior concentration that traditional methods would not choose as evaluation points (cf.\ \Cref{fig:helmholtz_results}, bottom left).
While many compact-domain measure transport maps
use bijections between $[0,1]^d$ and $\RR^d$ and then perform approximation on $\RR^d$~\cite{zengBoundedKRnet2025}, the concentration of this posterior near the boundary $\partial \Omega$ is particularly ill-suited to such approximations. In \Cref{app:map_param} we describe a class of transport maps $\FF$ that ``natively'' map from $[0,1]^d$ to itself; this parameterization ensures that the ridges of high probability, shown in \Cref{fig:helmholtz_results}, are adequately captured across the entire domain, especially near the domain boundaries. %

\begin{table}[ht!]
	\centering
	\caption{Hyperparameters and error metrics for \Cref{sec:helmholtz}. Row $\stepIDX$ corresponds to $Q_{\stepIDX}^\MIS$ and $\surrogateStep$. $n_\stepIDX = 25$ at each step.}%
	\begin{tabular}{@{}c *{2}{c@{\hskip 0.95\tabcolsep}} *{5}{c}@{}}
		\toprule	& \multicolumn{2}{c}{Hyperparameters} & & \multicolumn{4}{c}{Error Metrics}  \\
		\cmidrule(rl){2-3}  \cmidrule(rl){5-7}
		$\stepIDX$ & $\temperStep$ & $p_{\stepIDX}$ & rESS & RMSE & F\"{o}rstner & $\MMD_{\textrm{G}}$ & $\MMD_{\textrm{M}_{1.5}}$ \\
		\midrule
		0  & 0.075 & 14 & 8.10\E{-1} & 4.78\E{-1} & 5.38\E{-1} & 9.00\E{-1} & 9.04\E{-1} \\
		1  & 0.150 & 20 & 6.63\E{-1} & 4.71\E{-1} & 3.91\E{-1} & 7.77\E{-1} & 7.88\E{-1} \\
		2  & 0.300 & 20 & 5.37\E{-1} & 3.99\E{-1} & 3.00\E{-1} & 7.20\E{-1} & 7.34\E{-1} \\
		3  & 0.425 & 20 & 5.04\E{-1} & 3.12\E{-1} & 5.50\E{-1} & 7.07\E{-1} & 7.22\E{-1} \\
		4  & 0.450 & 27 & 5.00\E{-1} & 2.64\E{-1} & 4.78\E{-1} & 6.54\E{-1} & 6.70\E{-1} \\
		5  & 0.450 & 35 & 4.93\E{-1} & 3.32\E{-1} & 3.93\E{-1} & 6.41\E{-1} & 6.56\E{-1} \\
		6  & 0.500 & 44 & 5.01\E{-1} & 3.47\E{-1} & 3.85\E{-1} & 5.89\E{-1} & 6.00\E{-1} \\
		7  & 0.550 & 44 & 5.05\E{-1} & 5.42\E{-1} & 3.98\E{-1} & 6.13\E{-1} & 6.22\E{-1} \\
		8  & 0.625 & 54 & 5.04\E{-1} & 4.72\E{-1} & 3.50\E{-1} & 5.82\E{-1} & 5.91\E{-1} \\
		9  & 0.775 & 54 & 5.02\E{-1} & 5.18\E{-1} & 2.98\E{-1} & 5.55\E{-1} & 5.64\E{-1} \\
		10 & 0.825 & 65 & 5.02\E{-1} & 4.63\E{-1} & 2.80\E{-1} & 5.61\E{-1} & 5.69\E{-1} \\
		11 & 0.850 & 65 & 5.02\E{-1} & 5.87\E{-1} & 2.75\E{-1} & 5.69\E{-1} & 5.78\E{-1} \\
		12 & 0.875 & 65 & 5.00\E{-1} & 6.36\E{-1} & 2.81\E{-1} & 5.62\E{-1} & 5.69\E{-1} \\
		13 & 0.875 & 65 & 4.79\E{-1} & 6.85\E{-1} & 3.52\E{-1} & 5.65\E{-1} & 5.71\E{-1} \\
		14 & 1.000 & 77 & 5.03\E{-1} & 6.66\E{-1} & 3.08\E{-1} & 5.67\E{-1} & 5.73\E{-1} \\
		15 & 1.000 & 104 & 5.17\E{-1} & 6.41\E{-1} & 2.72\E{-1} & 5.75\E{-1} & 5.81\E{-1} \\
		\bottomrule
	\end{tabular}
	\label{tab:helmholtz_params}
\end{table}

These results also demonstrate that, while the posterior geometry is not obtained exactly, our algorithms work well in the single-fidelity setting, with 400 evaluations of the PDE solver. \Cref{tab:helmholtz_smc_mcmc} compares the accuracy and cost of our final quadrature rule to the output of state-of-the-art SMC and interacting-chain MCMC algorithms~\cite{dauWasteFreeSequentialMonte2022,goodmanEnsembleSamplersAffine2010}. In this comparison, we interpret $Q^{\MIS}_M$ as an empirical measure, gather an equivalently sized quadrature rule from each of these alternative algorithms (requiring more than 400 likelihood evaluations), and compute the MMD between the empirical measures and $\poster_M$. See \Cref{sm:helmholtz} for details. We find our $Q_M^\MIS$ yields smaller MMD at smaller cost. %

\begin{table}
	\centering
	\caption{Other methods' performance approximating $\poster_M$ in \Cref{sec:helmholtz}.}
	\label{tab:helmholtz_smc_mcmc}
	\begin{tabular}{r c c c}\toprule
		& \# of likelihood evaluations & $\MMD_{M_{1.5}}$ & $\MMD_{G}$\\\midrule
		$\boldsymbol{Q}_{\boldsymbol{M}}^{\textbf{MIS}}$ & 400 & 0.580 & 0.610\\
		SMC \cite{dauWasteFreeSequentialMonte2022} & 832 & 1.276 & 1.376\\
		MCMC \cite{goodmanEnsembleSamplersAffine2010} & 743 & 0.843 & 0.858\\
		\bottomrule
	\end{tabular}
\end{table}

We speculate that this performance could be improved further.
The number of PDE evaluations was fixed arbitrarily, and the function class $\FF_\stepIDX$ and other hyperparameters at step $\stepIDX$ were tuned in an ad hoc manner according to the a priori rESS diagnostic. %
We could also employ the multi-fidelity scheme, but the aim of this example is to highlight the standalone value of tempering and quadrature while comparing with related algorithms. %

\section{Conclusion}

We have developed an iterative framework for the approximation and sampling of geometrically complex target distributions, based on a generalized annealing process, variational construction of transport maps, and the adaptive combination of importance-weighted quadrature rules. Our algorithms are suited to computationally expensive black-box models and can exploit a collection of forward models of varying fidelity.
The aim of our framework is to \textit{meet Bayesian practitioners where they are}:  stranded without gradients and starved for model evaluations. Empirically, our generalized annealing scheme achieves good approximations of strongly non-Gaussian distributions with 100--400 samples, while enjoying significant parallelizability---at a sample budget where a user of MCMC or SMC would still be awaiting convergence.

There are many important directions in which to expand and improve the present methods.
First, our focus here has been on low-dimensional posterior distributions with expensive likelihoods and complex geometry. To extend the present approach to higher parameter dimensions, the modular structure of \Cref{alg:core_loop} need not change significantly, but the transport maps and quadrature rules considered therein should be adapted to exploit more structure in the prior and posterior. For instance, it is known that triangular transport maps inherit sparse variable dependence~\cite{tongMALAwithinGibbsSamplersHighDimensional2020} from conditional independence in the target~\cite{spantiniInferenceLowDimensionalCouplings2018,baptistaRepresentationLearningMonotone2023}. Similarly, many dimension reduction techniques for Bayesian inverse problems focus approximation and sampling on a low-dimensional subspace where the posterior departs most strongly from the prior. While many of these techniques are gradient-based~\cite{zahmCertifiedDimensionReduction2022,liSharpDetectionLowdimensional2025}, recent methods~\cite{baptista2025dimension} promise to identify such subspaces without gradient information and might be adapted here.
In higher dimensions, %
the geometric tempering used here might induce $Q_j$ with significantly lower rESS~\cite{chehab2025provable}. Thus, it is also of interest to consider other paths $(\pi_j)_j$, e.g., paths induced by Gaussian convolutions~\cite{zangerSequentialTransportMaps2024}. The design of ``good'' tempering paths, with notions of quality often quite specific to certain algorithms, is a very active area of research.

There is also a need for theory to characterize the convergence of our multiple importance-weighted quadrature schemes, given the cross-iteration dependencies and the use of structured quadrature rules (e.g., rQMC) as building blocks; this theory could yield sharper guidance on how to combine multiple rules, beyond the power heuristic and partition of unity used here.
On this note, we find that several of our final rules $Q_M^\MIS$ contain quadrature points close in space with vastly different weights. Quantization~\cite{belhadjiWeightedQuantization2025} or kernel thinning~\cite{dwivediKernelThinning2024} techniques could simplify these point sets to improve efficiency.

\section*{Acknowledgments}
DS would like to thank Mathieu Le Provost for particularly illuminating discussions about compact-domain transport. DS and BvBW acknowledge support from the Laboratory Directed Research and Development program at Sandia National Laboratories, a multimission laboratory managed and operated by National Technology
and Engineering Solutions of Sandia LLC, a wholly owned subsidiary of Honeywell International Inc., for the U.S.\ Department of Energy's National Nuclear Security Administration under contract DE-NA0003525. DS and YMM also acknowledge support from the U.S.\ Department of Energy, Office of Science, Office of Advanced Scientific Computing Research, Scientific Discovery through Advanced Computing (SciDAC) Program through the FASTMath Institute, under contract number DE-AC52-07NA27344. DS also acknowledges support from a 2025--26 MathWorks Engineering Fellowship at MIT. SAND2026-18510R.

\bibliographystyle{siamplain}
\bibliography{references}

\appendix
\section{Integral discretization details}
\label{app:qmc} Since we assume uniform prior $\pi_{\theta}=U{(0,1)}^{d}$, we use randomized QMC methods. More specifically, given a dimension $d$ and desired quadrature
evaluations $n$, we find the smallest prime base $b\in\mathbb{N}$ such that
$b > d$ and smallest padding $p\in\mathbb{N}$ such that $b^{p}> n$. Then, we take $n$ of the first $b^{p}$
points from a scrambled, $d$-dimensional QMC sequence. In \Cref{sec:examples}, we use the Sobol' sequence~\cite{sobolDistributionPointsCube1967} and employ
Owen's scramble~\cite{owenRandomlyPermutedTmsNets1995}, both implemented in~\cite{variouscontributorsSciMLQuasiMonteCarlojl2019}.

\section{Adaptive selection of tempering parameter}
\label{app:temp_sched} We briefly summarize an adaptive method inspired by~\cite{liAdaptiveConstructionSurrogates2014,wangMitigatingModeCollapse2025},
adjusted for our use-case, to choose tempering parameter $\beta$ adaptively. Recall
the definition of the relative effective sample size $\rESS$ for a normalized set of weights ${(w^{(k)})}
_{k=1}^{N}\subset[0,1]$ as $\rESS[\mathbf{w}] = (N\|\mathbf{w}\|_2^{2})^{-1}.$
Then, we fix a global discount parameter $\rho$ and minimum $\rESS$ value $r_{\mathrm{min}}\in (0,1)$.
We choose $\rho = 0.8$ and $r_{\mathrm{min}}= 0.5$ and create a monotone sequence of tempering candidates $\mathcal{B} = (b_q)_{q=1}^{N_\beta}\subset (0,1 )$, chosen as $b_q = q/N_\beta$ for $N_\beta = 40$ in \Cref{sec:examples}. Then, our procedure
to choose $\beta_{j+1}$ is given in \Cref{alg:temp_sched}.

\begin{algorithm}
	[ht!]
	\caption{Adaptively choosing tempering parameter}%
	\label{alg:temp_sched}
	\begin{algorithmic}
		\REQUIRE{Tempering candidates $\mathcal{B}$, discount $\rho \in (0,1)$, previous $\rESS$ value $r_{j}$, previous tempering $\beta_{j}$, $\rESS$ floor $r_{\mathrm{min}}\in(0,1)$, and \Cref{alg:multiple_IS} arguments $A={(\surrogate_{i-1},D_i)}_{i=j_0}^{j}$}%
		\ENSURE{Tempering parameter $\beta_{j+1}$, quadrature rule $Q_{j+1}$, and $\rESS$ $r_{j+1}$}
		\STATE{\textbf{set} index $q = N_{\beta}$ and $r_{\mathrm{min}}= \max(\rho\,r_{j},r_{\mathrm{min}})$}
        \WHILE{$q > 0$ and $r_q > r_{\mathrm{min}}$ and $b_{q} \leq \beta_j$}
            \STATE{Create quadrature rule $P_{q}$ from \textbf{\Cref{alg:multiple_IS}} with tempering $\beta = b_{q}$ and arguments $A$.}
			\STATE{\textbf{set} $r_{q}= \rESS[P_{q}]$ then decrement $q$.}
        \ENDWHILE
		\STATE{\textbf{set} $\beta_{j+1}= b_{q+1}$, $Q_{j+1}= P_{q+1}$, and $r_{j+1}= \rESS[P_{q+1}]$}
	\end{algorithmic}
\end{algorithm}
We use \Cref{alg:temp_sched} \textit{in place} of \Cref{alg:multiple_IS} inside of
\Cref{alg:core_loop}\todofinal{NOTE}. %
In future work, we propose varying minimum threshold $r_{\mathrm{min}}$ as $\beta_{j}$
increases; while \Cref{alg:temp_sched} allows for some flexibility above $r_{\mathrm{min}}$, we may want to increase threshold $r_{\mathrm{min}}$ as $j\to M$, i.e., as $\beta_{j}\to 1$.

\section{Parameterizing transport maps on the unit hypercube}
\label{app:map_param} The Knothe--Rosenblatt rearrangement (often also called KR map or triangular transport
map) is a popular measure transport method~\cite{knotheContributionsTheoryConvex1957,rosenblattRemarksMultivariateTransformation1952,villaniOptimalTransportOld2009,baptistaRepresentationLearningMonotone2023,parnoTransportMapAccelerated2018},
which in one-dimension is exactly the ``inverse-CDF trick''~\cite{devroyeNonUniformRandomVariate1986}:
if we have $Z\sim \eta=U(0,1)$ and a target density $\pi\in\mathcal{C}^{1}$ with cumulative distribution
function (CDF) $F_{\pi}:\RR\to[0,1]$, we know that $F_{\pi}$ is invertible and $F_{\pi}^{-1}(Z)\sim\pi$.
We thus have identity ${F_\pi}_\sharp\pi = \eta$ and its converse $F_\pi^\sharp\eta = \pi$.
For the uniform reference case, i.e., when $\eta=U{(0,1)}^{d}$, the KR map is an extension of the
above by factoring the joint density into its conditionals:
\[
	\pi(\theta_{1},\theta_{2},\ldots,\theta_{d}) = \pi^{(1)}(\theta_{1})\;\pi^{(2)}(\theta_{2}|\theta_{1}
	)\;\pi^{(3)}(\theta_{3}|\theta_{1},\theta_{2})\;\cdots\;\pi^{(d)}(\theta_{d}|\theta_{1},\ldots,\theta
	_{d-1}).
\]
Given $Z_{1},\ldots,Z_{d}\sim U(0,1)$, we set $\theta_{1}= F_{\pi^{(1)}}^{-1}(Z_{1})$ and know
$\theta_{1}\sim \pi^{(1)}$. Conditioned on this fixed $\theta_{1}$, univariate density $\pi^{(2)}$
has CDF
$F_{\pi^{(2)}}(\theta_{2};\theta_{1})=\int_{0}^{\theta_2}d\pi^{(2)}(t|\theta_{1}) \ dt$, so $\theta_{2}
=F_{\pi^{(2)}}{(\cdot;\theta_1)}^{-1}(Z_{2})\sim\pi^{(2)}(\theta_{2}|\theta_{1})$; this continues
for $j=3,\ldots,d$ via
\begin{equation}\label{eqn:kr_conditional_cdf}
	\theta_{j}= F_{\pi^{(j)}}^{-1}(\cdot\ ;\theta_{1},\ldots,\theta_{j-1})(Z_{j}), \quad F_{\pi^{(j)}}(
	\theta_{j};\theta_{1},\ldots,\theta_{j-1}) = \int_{0}^{\theta_j}\pi^{(j)}(t\ |\ \theta_{1},\ldots,\theta
	_{j-1})\ dt.
\end{equation}
This transport map construction has earned the ``triangular transport'' moniker since the map
$S=(F_{\pi^{(1)}},F_{\pi^{(2)}},\ldots,F_{\pi^{(d)}})$ has a lower-triangular Jacobian matrix, an artifact
of its conditional CDF construction, meaning that the pullback density
$\pi(\theta)=\eta(S(\theta))|\nabla S(\theta)|$ is simple to calculate. %
There are many approximations of KR maps in literature, e.g.,~\cite{sargsyanSpectralRepresentationReduced2010,wangEfficientNeuralNetwork2024,cuiScalableConditionalDeep2023,zangerSequentialTransportMaps2024,ecksteinComputationalMethodsAdapted2024}.
We choose a parametric approximation from~\cite{wangMinimaxDensityEstimation2022}, %
which has interesting theoretical results but no prior implementation, though
it is heavily inspired work in, e.g.,~\cite{marzoukSamplingMeasureTransport2016,baptistaRepresentationLearningMonotone2023,parnoTransportMapAccelerated2018}.
Using reference $\eta=U(0,1)^d$ and recalling the conditional CDF structure of $S$ from~\eqref{eqn:kr_conditional_cdf}, a natural characterization of $S^{(d)}$ is to then create a \textit{positive-valued} parametric function
$f_{d}(\cdot,\cdot;\bfc_{d}):{[0,1]}^{d-1}\times[0,1]\times\RR^{p_d}\to\RR^{+}$, parameterized by some
$\bfc_{d}\in\RR^{p_d}$, and use $f_{d}$ to proportionally approximate
$\pi^{(d)}(t|\theta_{1:d-1})\propto f_{d}(\theta_{1:d-1},t;\bfc_{d})$. Then, %
\[
	I_{d}(s;\theta_{1:d-1},\bfc_{d}):=\int_{0}^{s}f_{d}(\theta_{1:d-1},t;\bfc_{d}) \ dt,\quad S^{(d)}(\theta;c_d) := \frac{I_{d}(\theta_{d};\theta_{1:d-1},\bfc_{d})}{I_{d}(1;\theta_{1:d-1},\bfc_{d})}.
\]

Since $f_{d}\approx \pi^{(d)}$ is an unnormalized conditional density, the denominator ensures $S^{(d)}(\theta_{1:d-1},1) \equiv 1$. Therefore, we calculate the CDF of the
conditional distribution $\pi^{(d)}$ by employing its unnormalized approximation and then
normalizing it. Notably, both integrals in this expression are one-dimensional and thus tractable
using traditional quadrature techniques. We do not vary inputs $\theta_{1:d-1}$ or $\bfc_{d}$ while calculating
$I_{d}$, indicating that the parameterization of $f_{d}$ dictates the expense to evaluate
$I_{d}$. %
We choose to use $f_{d}(\theta_{1:d-1},t;\bfc_{d}) = r(g_{M}{(\theta_{1:d-1},t)}^{\top}
\bfc_{d})$, where rectifier $r:\RR\to\RR^+$ is a positive, monotone increasing function, and $g_{M}:\mathbb{R}^{d}\to\mathbb{R}^{p_d}$ is an arbitrary set of basis functions. In this paper, we use rectifier $r(t) = \log(1+\exp(t))$ and $g_M$ as a total order $M$ set of tensor-product
shifted Legendre polynomials. Each output of $g_{M}$ is thus a product of univariate
shifted Legendre polynomials and has at most $M+1$ nonzero mixed derivatives, i.e.,
for all $a\in [p_{j}]$, we know $\frac{\partial^{\vec{\alpha}}}{\partial \theta^{\vec{\alpha}}}g_{a}
= 0$ when multi-index $\vec{\alpha}$ satisfies $\|\vec{\alpha}\|_{1}> M$. For more on triangular transport with details on our implementation and optimization, see \Cref{sm:expansions}.
Further, see~\cite{baptistaRepresentationLearningMonotone2023,bigoniAdaptiveConstructionMeasure2016,marzoukSamplingMeasureTransport2016}
for computation and~\cite{ramgraberFriendlyIntroductionTriangular2025} for a
tutorial.

\section{Example of multiple importance sampling}
We use a toy set of parameters to illuminate the complex connections of multiple IS, particularly with
a focus on the cumulative integration technique proposed in \Cref{alg:multiple_IS}. Suppose
there is a point $\theta_{0}^{*}$ such that $\prior(\theta^{*}_{0})= 0.25$ but
$\surrogate_{1}(\theta^{*}_{0}) = 0.75$ with $n_{0}= 2n_{1}$. Then,
\[
	\alpha_{0}(\theta^{*}_{0}) = \frac{{(n_0\prior(\theta^*_0))}^{2}}{{(n_0\prior(\theta^*_0))}^{2}+{(n_1\surrogate_0(\theta^*_0))}^{2}}
	= \frac{{(0.25)}^{2}}{{(0.25)}^{2}+{(1.5)}^{2}}\approx 0.027,
\]
exemplifying that, if $\widehat\poster_{j}\ll \widehat\poster_{j^\prime}$ at $\theta^{(k)}_{j}$, then
$\alpha_{j}(\theta^{(k)}_{j})$ will be very small. On the other hand, if the two biasing
distributions give similar values, the integral is split more equally between the biasing distributions
near $\theta^{(k)}_{j}$.

\section{Further discussion for transport maps}
\label{sm:expansions}
Below, we include a few supplemental remarks detailing the relationship of triangular maps used in this paper to optimal transport, often important for readers unfamiliar with statistical transport map techniques. We also include details of the parameterization, which is largely an extension of discussions in~\cite{ramgraberFriendlyIntroductionTriangular2025}, and add a few implementation notes vital for reproducing results.

\subsection{Relating KR maps and Optimal Transport}
While the one-dimensional case boils down to the exact same solution as one would get performing \textit{optimal
transport}, the KR map generally will not coincide with the ``optimal'' map as the reader may see discussed
in, e.g.,~\cite{feydyInterpolatingOptimalTransport2019,villaniOptimalTransportOld2009}; instead, it
reflects some sense of ``conditional optimality'' because of the sequence of conditional CDF
transports~\cite{hosseiniConditionalOptimalTransport2024,wangEfficientNeuralNetwork2024,bonnotteKnothesRearrangementBreniers2013}.
Regardless, the lack of optimality will not change the algorithm nor will it disrupt the
practicality, for we only seek to couple $\poster$ and $\refer$ via some map $S$. We have no need for $S$
to perform minimal work so long as it takes us between $\poster$ and $\refer$ in an invertible fashion, and
the triangular map gives a computationally convenient methodology for doing so (much more so than
most optimal transport methods).

\subsection{Details of polynomial expansions}
Consider a set of univariate bases ${\{\psi^{(d)}_k\}}_{d,k}$, with $\psi^{(d)}_{k}$ as the $k$th univariate
basis function for input dimension $d$. For example, consider a two-dimensional problem with $\psi^{(1)}
_{k}(\theta)=\sin (2\poster k \theta)$ and $\psi^{(2)}_{k^\prime}$ as the $k^{\prime}$th Legendre
polynomial $\widetilde{p}_{k^\prime}$. These are both orthogonal sets on $L_{2}([ 0,1])$. Similar to
multivariate polynomial surrogate construction~\cite{jakemanPolynomialChaosExpansions2019,xiuWienerAskeyPolynomial2002,conradAdaptiveSmolyakPseudospectral2013},
we choose a set of multi-indices $\mathcal{A}\subset\mathbb{N}_{0}^{d}$ to describe the function class
for $f_{d}$; in all examples, we choose \textit{total order} sets
$\mathcal{A}^{d}_{T}:= \{\vec{\alpha}\in\mathbb{N}_{0}^{d}\ |\ \|\vec{\alpha}\|_{1}\leq T\}$. We understand
$\vec{\alpha}$ as representing a \textit{multivariate basis function}
$\Psi_{\vec{\alpha}}(\theta) = \prod_{j=1}^{d}\psi^{(j)}_{\alpha_{\stepIDX}}(\theta_{j})$. This inundates the
reader with notation, but returning to the prior example, multi-index $\vec{\alpha}=(2,1)$ produces multivariate
basis function $\Psi_{(2,1)}= \sin(4\poster \theta_{1})\widetilde{p}_{1}(\theta_{2})$. We then create an
expansion
$\widehat{f}_{d}(\theta;\bfc_{d}) = \sum_{\vec{\alpha}\in\mathcal{A}}c_{\vec{\alpha}}\Psi_{\vec{\alpha}}$.
Further, to satisfy the positivity constraint $f_{d}> 0$, we choose a \textit{rectifier}
$r:\RR\to\RR^{+}$ (e.g., $r(s)=\exp(s)$ or $r(s)=\log(1+\exp(s))$). Finally, we set %
$f_{d}(\theta_{1:d};\bfc_{d})=r(\partial_{d}\widehat{f}_{d}(\theta_{1:d};\bfc_{d}))$. The intuition
here in the scalar case is that, for rectifier $r(s)=\exp(s)$ and reference density $\refer=U(0,1)$,
we get
$\log S^{\sharp}\refer = \log\refer\circ S + \log \partial_{\theta}S = \widehat{f}(\theta ;\bfc) - \log\int
_{0}^{1}\exp(\widehat{f}(t;\bfc))\ dt$, so we explicitly perform parametric density estimation via $\widehat
{f}\approx\log\poster+C$

\subsection{Optimization and hyperparameters}
All approximation for the transport maps in \Cref{sec:examples} are shifted Legendre polynomials
with the softplus rectifier and adaptive Clenshaw-Curtis quadrature as implemented in~\cite{parnoMParTMonotoneParameterization2022}. %
For optimizing $S$, we use $10^3$ Nesterov optimization
steps (with default hyperparameters of $\eta = 0.001, \rho = 0.9$) of size $10^{-3}$ as implemented in~\cite{variouscontributorsFluxMLOptimisersjl2022},
with $L_2$ regularization of size $10^{-3}$. For total order maps, we have $p_{d}=\mathrm{nCr}(M+d, d)$ with binomial coefficient $\mathrm{nCr}$. \Cref{tab:ex_poly_orders} thus describes the total order used for each example in \Cref{sec:examples}.

\begin{table}[ht!]
	\centering
	\caption{Total order of transport maps for each step of examples. Bold indicates the first step for a given likelihood fidelity.}%
	\label{tab:ex_poly_orders}
	\begin{tabular}{@{}*{17}{c}@{}}
		\toprule                                                          & \multicolumn{16}{c}{$j$} \\
		\cmidrule(l){2-17} Example &
		0 & 1 & 2 & 3 & 4 & 5 & 6 & 7 & 8 & 9 & 10 & 11 & 12 & 13 & 14 & 15 \\
		\midrule \hyperref[sec:diffusion_single]{Single source diffusion} &
		\textbf{4} & 4 & 5 & 5 & \textbf{3} & \textbf{3} & 7 & \multicolumn{9}{l}{\cellcolor{gray!60}}\\
		\hyperref[sec:diffusion_multi]{Multiple source diffusion} &
		\textbf{4} & 6 & 6 & \textbf{3} & \textbf{3} & 6 & 7 &
		\multicolumn{9}{l}{\cellcolor{gray!60}} \\
		\hyperref[sec:cdr]{Convection diffusion reaction} &
		\textbf{3} & 5 & \textbf{4} & 5 & \textbf{3} & 5 & 7 & 7 & 7 &
		\multicolumn{7}{c}{\cellcolor{gray!60}} \\
		\hyperref[sec:helmholtz]{Helmholtz} &
		\textbf{4} & 5 & 5 & 5 & 6 & 7 & 8 & 8 & 9 & 9 & 10 & 10 & 11 & 12 & 12 & 13 \\
		\bottomrule
	\end{tabular}
\end{table}%

\section{Error metric background and details}
\label{sm:error_metrics} First, given a symmetric-positive kernel function
$K:\RR^{d}\times\RR^{d}\to\RR^{+}$ (i.e., $K(x,y) = K(y,x) \geq 0$), we define the MMD as
\[
	\mathrm{MMD}_{K}^{2}(\poster,\surrogate) = \EE_{\poster,\poster}[K(X,X^{\prime})] -2\EE_{\poster,\surrogate}
	[K(X,Y)] + \EE_{\surrogate,\surrogate}[ K(Y,Y^{\prime})].
\]%
We use two distance-based kernels: squared-exponential and Mat\`{e}rn-1.5, denoted
$K_{S}(x,y) = k_{S}(\|x-y\|_{2}/\sigma_{\mathrm{MMD}})$ and $K_{M_{1.5}}(x,y) = k^{1.5}_{M}(\|x- y\|_{2}/\sigma_{\mathrm{MMD}})$, respectively,
with tunable bandwidth hyperparameter $\sigma_{\mathrm{MMD}}$ defining the length-scale of a feature. We then define the
squared-exponential and Mat\`{e}rn kernel functions respectively as
\[
	k_{S}(t) = \exp(-t^{2}/2),\quad k_{M}^{\nu}(t) = \frac{2^{1-\nu}}{\Gamma(\nu)}{\left(\sqrt{2\nu}t\right)}
	^{\nu}\kappa_{\nu}{\left(\sqrt{2\nu}t\right)},
\]
with $\kappa_{\nu}$ as the modified $\nu$-th Bessel's function of the second kind, where we use
$\nu=1.5$ for this article. %
As it is infinitely differentiable, the $\mathrm{MMD}_{K_S}$ tends to be able to only identify broader
features and ``smooths out'' distances, which makes it good for identifying when we want to check
for the correct broad behavior. On the other hand, kernel function $k^{\nu}_{M}$ is only once
differentiable for $\nu=1.5$ at $0$ and thus $\mathrm{MMD}_{K_{M_{1.5}}}$ tends to reflect how well $\poster$
and $\surrogate$ match higher-frequency features~\cite{muandetKernelMeanEmbedding2017}. We also use the simpler error metrics
of
\[
	d^{2}_{\mathrm{RMSE}}(\poster,\surrogate) = \|\EE_{\poster}[\theta] - \EE_{\surrogate}[\theta]\|^{2}_{2}
	,\quad d_{F}^{2}(\poster,\surrogate) = \sum_{i=1}^{d}\ln^{2}\lambda_{i}\left(\mathbb{C}\mathrm{ov}[\poster
	],\mathbb{C}\mathrm{ov}[\surrogate]\right),
\]
where $\mathbb{C}\mathrm{ov}[\mu]$ is the covariance matrix of pdf $\mu$ and $\lambda_{i}(\mathbf{A},
\mathbf{B})$ is the $i$th-largest generalized eigenvalue of $\mathbf{A}$ and $\mathbf{B}$. These are
the RMSE and F\"{o}rstner~\cite{forstnerMetricCovarianceMatrices2003} distances. All discrepancies described
are nominally semi-metrics (i.e., zero error in all of these is necessary but not sufficient for
$\poster\equiv \surrogate$). For each example, we report each error relative the prior, e.g.,
for posterior $\poster_{M}$ and $j$-th approximation $\surrogate_{j}$, we report $\widetilde{D}(\poster_{M}
,\surrogate_{j}) = D(\poster_{M},\surrogate_{j})/ D(\poster_{M},\prior)$ for distance $D$ as each
of the RMSE, F\"{o}rstner, $\mathrm{MMD}_{K_S}$ and $\mathrm{MMD}_{K_{M_{1.5}}}$ discrepancies. Since we
learn $\surrogate$ via measure transport, this is intended to non-dimensionalize the errors
relative to ``how much work we would have to do to make $d$ zero going from prior to posterior.'' In
other words, if $D(\poster,\prior) = \eps > 0$, then we certainly would require $D(\poster,\surrogate) < \eps$ to consider the method successful.

Notably, all of these require us to integrate over both $\surrogate$ and $\poster$. For $D_{\mathrm{RMSE}}$
and $D_{F}$, we measure the error between the moment of the true $\poster_{M}$ and the estimated moments
from normalized importance samples at step $j$, i.e., ${(\theta^{(k)}_{\stepIDX},w^{(k)}_{\stepIDX})}_{k=1}^{N_{\stepIDX}}$,
using covariance estimation formula%
\[
	\mathbb{C}\mathrm{ov}[\poster_{j}]\approx \frac{1}{1-\sum\limits_{k=1}^{N}{(w_{\stepIDX}^{(k)})}^{2}}\sum_{k=1}^{N}
	w_{j}^{(k)}(\theta_{j}^{(k)}- \widehat{\mu}_{j}){(\theta_{\stepIDX}^{(k)} - \widehat{\mu}_{\stepIDX})}^{\top},\quad \widehat
	{\mu}_{j}= \sum_{k=1}^{N}w_{j}^{(k)}\theta_{j}^{(k)}.
\]
To integrate with respect to a given unnormalized posterior density $\poster= \poster^{\bfy^*}\prior
\propto\poster$, we choose a fine-grid quadrature $(\theta_{\poster}^{(k)},\widetilde{w}_{\poster}^{(k)})_{k=1}^N$
on $\prior$ and then reweight according to
$w_{\poster}^{(k)}\propto \poster^{\bfy^*}(\theta^{(k)}_{\poster})\widetilde{w}_{\poster}^{(k)}$, where the proportionality
comes from normalization of $\sum_{k}w_{\poster}^{(k)}\equiv 1$. For transport based
$\surrogate= S^{\sharp}\refer$, we calculate $\mathrm{MMD}$-based errors via a fine-grid pullback of
quadrature rule $(z_{\refer}^{(k)},\ w_{\refer}^{(k)})_{k=1}^N$ on $\refer$ via $\theta_{\refer}^{(k)}= S^{-1}(z_{\refer}
^{(k)})$. %

\section{Setup for numerical experiments}
\label{sm:fem_discretization}
For all experiments, we use continuous Galerkin discretizations with structured uniform meshes. For all two-dimensional examples the $\MMD$, RMSE, and F\"{o}rstner calculations use an order $50$ tensorized Gauss--Legendre quadrature rule (i.e., requiring 2500 evaluations of $\poster$), where the $\MMD$ bandwidth is $\sigma_{\mathrm{MMD}}=0.05$. Below are specific choices for each example.

\paragraph{Diffusion PDE single source inversion}
We use first-order triangular elements with second-order quadrature for all solvers, where the data is generated with $256^2$ elements. Likelihood $\likelyFidel$ has $(2m_\fidelity)^2$ elements for $m_{0}=8$, $m_1 = 32$, and $m_2=64$.

\paragraph{Diffusion PDE multi-source inversion}
We use $256^2$ second-order triangular elements with fourth-order quadrature for the data generation. Likelihood $\likelyFidel$ uses $(2m_\fidelity)^2$ first-order elements with second-order quadrature, where $m_0=8$, $m_1=16$, and $m_2=64$.

\paragraph{Convection diffusion reaction}
We semi-discretize using first-order triangular elements with second-order quadrature for all solvers, where the data is generated with $256^2$ elements and convection speed $c^*=0.5$. All likelihoods $\likelyFidel$ use $32^2$ elements, with differing convection speed $c_\fidelity$, where $c_0=0$, $c_1=0.25$, and $c_2=c^*$. These semi-discretizations are all solved using backward Euler on $t\in[0,10]$ with $\Delta t=0.1$ using a first-order BDF implicit solver. All $\MMD$ calculations use a bandwidth $\sigma=0.1$ and the first $d^{7} = 16,384$ four-dimensional Sobol' points for the quadrature.
The KDEs in \Cref{fig:cdr_densities,fig:cdr_pullback} are estimated using these $16384$ Sobol' points with a squared-exponential
kernel, bandwidth calibrated according to $\sigma_{\mathrm{KDE}} N^{-1/6}$, and $\sigma_{\mathrm{KDE}}$ as the average marginal
standard deviation across all dimensions. For each subplot in \Cref{fig:cdr_densities,fig:cdr_pullback}, the contour colors are calibrated as
identical and the vertical scaling for the KDEs on the diagonal are identical.

\paragraph{Helmholtz}
We use quadrilateral second-order elements with fourth-order quadrature for the solvers, where the data is generated using $128^2$ elements and the (single) likelihood uses $64^2$ elements.

\section{Numerical comparisons}\label{sm:helmholtz}
Here we describe and justify our comparisons in \Cref{sec:examples}. In \Cref{sec:diffusion_multi}, we suggest that the Monte Carlo standard error for the mean would be approximately $3.42\times 10^{-2}$ when using $N=75$ samples. This is derived using the $50^2$-point Gauss--Legendre quadrature rule to estimate the variance $\sigma_j$ of each marginal $j$. We then recall the Lindeberg-L\'{e}vy central limit theorem: if $\overline{\Theta}$ is a sample mean of random variable $\Theta$ with mean $\mu$ and variance $\sigma^2$, then the law of $\sqrt{n}(\overline{\Theta} - \mu)$ converges to a centered normal distribution with variance $\sigma^2$. Using the Monte Carlo standard error of $\sigma / \sqrt{N}$, then, gives the $3.42\times 10^{-2}$ for each marginal (where each marginal has $\sigma \approx 0.296$).

In \Cref{sec:helmholtz}, we compare our method to SMC~\cite{dauWasteFreeSequentialMonte2022} and MCMC~\cite{goodmanEnsembleSamplersAffine2010}. The first algorithm, Waste-Free SMC, is built on a sequence of tempered densities on which one performs Metropolis-like steps when performing SMC resampling. The ``waste-free'' nomenclature refers to the limiting of waste \textit{inside} the metropolization; the entire algorithm, however, does not necessarily make use of every evaluation in the final produced set of weighted samples. This explains why, in \Cref{tab:helmholtz_smc_mcmc}, the algorithm produces 400 samples but requires 832 evaluations. The implementation was provided by the \texttt{particles} package~\cite{chopinIntroductionSequentialMonte2020}, where we use the default adaptive tempering scheme with 25 particle ensembles and length 16 chains (the product of these two choices explains the size-400 rule). Similarly, the stretch MCMC algorithm~\cite{goodmanEnsembleSamplersAffine2010} produces an interacting set of ``walkers'' (each of which produces a Markov chain), but does not necessarily use tempering or simulated annealing by default. We use 20 walkers with length 120 chains in UQPy~\cite{olivierUQpyGeneralPurpose2020}, with a burn-in of 10 evaluations and thinning the chain by a factor of 10. For both SMC and MCMC, we only count evaluations as proposals that fall in the hypercube domain---i.e., if the algorithm proposed evaluating outside $[0,1]^2$, we do not count this toward the evaluation count in \Cref{tab:helmholtz_smc_mcmc}. Here, the rESS of SMC is 0.385 once the points are made uniquely (resulting in 79 points out of the original 400 samples); the rESS of MCMC of the original chains (before thinning) average about
\begin{equation}
	\mathbb{E}_{\mathrm{chains}}\left[\frac{1}{1 + 2\sum_{k=1}^\infty \rho_k}\right] \approx (0.03833, 0.04078),
\end{equation}
with an acceptance rate of 0.3121 averaged across the chains. Here, we estimate the length-$k$ autocorrelation $\rho_k$ for $0 \leq k\leq 30$, and each output corresponds to the rESS of each dimension.

\begin{figure}
	\includegraphics[width=\textwidth]{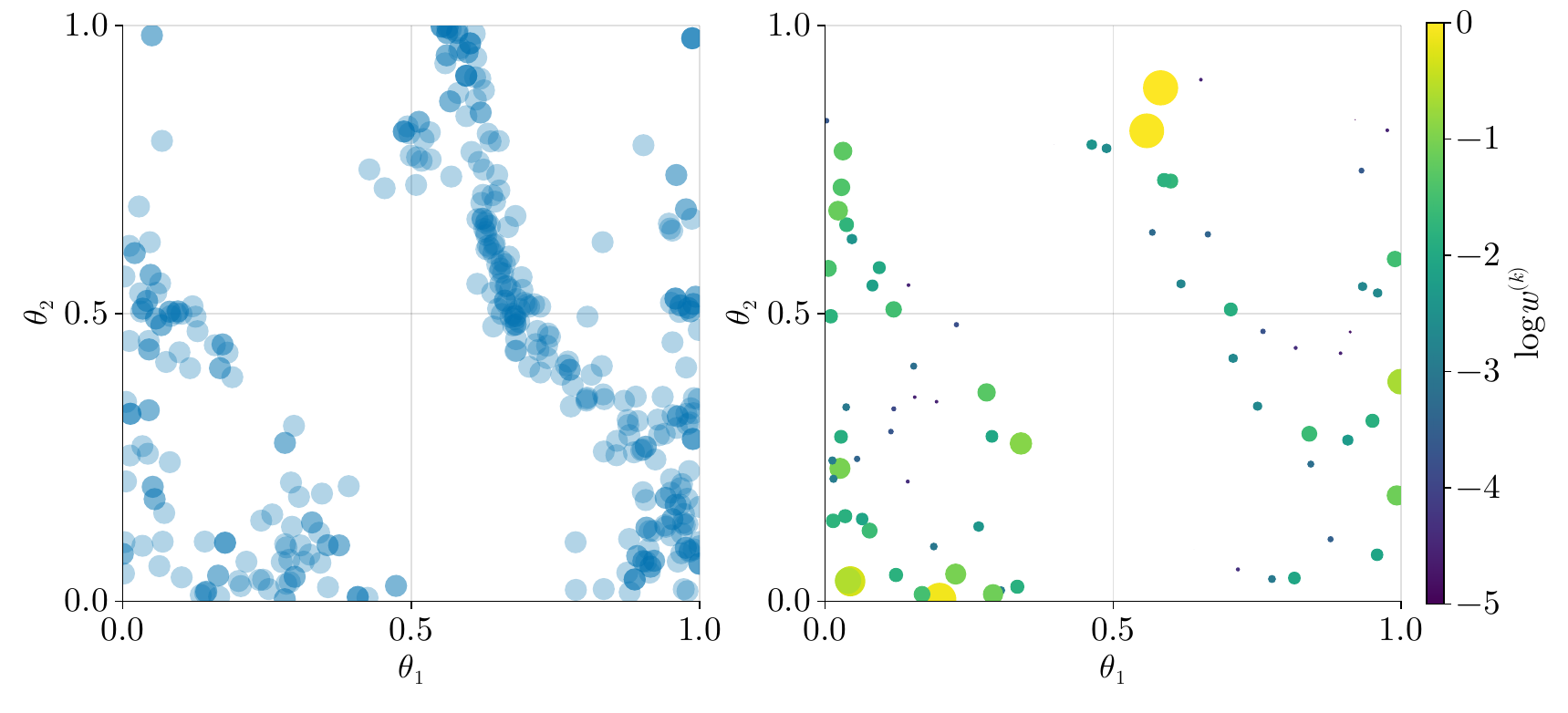}
	\caption{Output quadratures of MCMC (left) and SMC (right). SMC marker size and color reflect the weight and log-weight of the sample point, respectively; MCMC is shown with opacity to demonstrate the rejections in the Markov chain.}
	\label{sm_fig:helmholtz_comparison}
\end{figure}

We visualize the resulting quadratures of the SMC and MCMC in \Cref{sm_fig:helmholtz_comparison}. For reference, compare these to \Cref{fig:helmholtz_results}, top left. Finally, for comparing the quadrature rules to one another we use the \textit{sample-based} MMD calculation (with identical MMD bandwidth as in \Cref{sm:fem_discretization}). Given normalized quadrature rules $Q_1$ and $Q_2$, the sample-based MMD is calculated as

\begin{equation}
	\begin{gathered}
		\mathrm{MMD}^2(Q_1, Q_2) = \overline{K}(Q_1, Q_1) - 2 \overline{K}(Q_1, Q_2) + \overline{K}(Q_2, Q_2)\\
		\overline{K}(Q, U) = \sum_{k=1}^{N_Q}\sum_{k^\prime=1}^{N_U} K(\theta^{(k)},\nu^{(k^\prime)})w^{(k)}u^{(k^\prime)}\ \ \textrm{with}\ \ Q = (\theta^{(k)}, w^{(k)})_{k=1}^{N_Q},\ U = (\nu^{(k)}, v^{(k)})_{k=1}^{N_U}.
	\end{gathered}
\end{equation}

\end{document}